\documentstyle[10pt,epsf,epsfig,dp_delphititle,oldlfont]{dp_delphi}
\makeindex
\def\DpPaperGroup{EP}
\def\DpPaperRef{2001-002}
\def\DpDate{8 January 2001}
\def\DpAuthors{DELPHI Collaboration}
\def\DpSubmit{(Accepted by Phys.Lett.B)}
\def\DpTitle{\boldmath Measurement of $V_{cb}$ from the decay process
 \mbox{$\bar{B^{0}}\rightarrow D^{*+} \ell^- \bar{\nu}$}}
\def\DpComment{ }
\def\DpEMail{ }

\begin{document}
\makeatletter
\newcount\@tempcntc
\def\@citex[#1]#2{\if@filesw\immediate\write\@auxout{\string\citation{#2}}\fi
  \@tempcnta\z@\@tempcntb\m@ne\def\@citea{}\@cite{\@for\@citeb:=#2\do
    {\@ifundefined
       {b@\@citeb}{\@citeo\@tempcntb\m@ne\@citea\def\@citea{,}{\bf ?}\@warning
       {Citation `\@citeb' on page \thepage \space undefined}}%
    {\setbox\z@\hbox{\global\@tempcntc0\csname b@\@citeb\endcsname\relax}%
     \ifnum\@tempcntc=\z@ \@citeo\@tempcntb\m@ne
       \@citea\def\@citea{,}\hbox{\csname b@\@citeb\endcsname}%
     \else
      \advance\@tempcntb\@ne
      \ifnum\@tempcntb=\@tempcntc
      \else\advance\@tempcntb\m@ne\@citeo
      \@tempcnta\@tempcntc\@tempcntb\@tempcntc\fi\fi}}\@citeo}{#1}}
\def\@citeo{\ifnum\@tempcnta>\@tempcntb\else\@citea\def\@citea{,}%
  \ifnum\@tempcnta=\@tempcntb\the\@tempcnta\else
   {\advance\@tempcnta\@ne\ifnum\@tempcnta=\@tempcntb \else \def\@citea{--}\fi
    \advance\@tempcnta\m@ne\the\@tempcnta\@citea\the\@tempcntb}\fi\fi}
 
\makeatother
\begin{titlepage}
\pagenumbering{roman}
\CERNpreprint{\DpPaperGroup}{\DpPaperRef} 
\date{{\small\DpDate}} 
\title{\DpTitle} 
\address{\DpAuthors} 
\begin{shortabs} 
\noindent
%
A new precise measurement of $|V_{cb}|$ and of the branching ratio
\mbox{BR$(\bar{B^0} \rightarrow D^{*+} \ell^- \bar{\nu_\ell}$)}
has been performed using  a sample of about 5000 semileptonic decays 
\mbox{$\bar{B^0} \rightarrow D^{*+} \ell^- \bar{\nu_\ell} $}, 
selected by the DELPHI detector at LEP I 
by tagging the soft pion from \mbox{$D^{*+} \rightarrow D^0 \pi^+$}. The results are:
\begin{eqnarray}
\nonumber V_{cb} =(39.0 \pm 1.5 ~\mathrm{(stat.)}~ ^{+2.5}_{-2.6}
             ~\mathrm{(syst.~exp.)}~ \pm 1.3 ~\mathrm{(syst.~th.)})\times 10^{-3} \\
\nonumber {\mathrm{BR}}(\bar{B^0} \rightarrow D^{*+} \ell^- \bar{\nu_\ell}) = 
 (4.70 \pm 0.13~\mathrm{(stat.)}~ ^{+0.36}_{-0.31} ~\mathrm{(syst.~exp.)})\%
\end{eqnarray} 
The analytic dependences of the differential cross-section and of the Isgur 
Wise form factor as functions  of the variable $w = v_{B^0} \cdot v_{D^*}$
have also been obtained by unfolding the experimental resolution.


\end{shortabs}
\vfill
\begin{center}
\DpSubmit \ \\ 
\DpComment \ \\
\DpEMail \ \\
\end{center}
\vfill
\clearpage
\headsep 10.0pt
\addtolength{\textheight}{10mm}
\addtolength{\footskip}{-5mm}
\begingroup
%
\newcommand{\DpName}[2]{\hbox{#1$^{\ref{#2}}$},\hfill}
\newcommand{\DpNameTwo}[3]{\hbox{#1$^{\ref{#2},\ref{#3}}$},\hfill}
\newcommand{\DpNameThree}[4]{\hbox{#1$^{\ref{#2},\ref{#3},\ref{#4}}$},\hfill}
\newskip\Bigfill \Bigfill = 0pt plus 1000fill
\newcommand{\DpNameLast}[2]{\hbox{#1$^{\ref{#2}}$}\hspace{\Bigfill}}
%
\footnotesize
\noindent
\DpName{P.Abreu}{LIP}
\DpName{W.Adam}{VIENNA}
\DpName{T.Adye}{RAL}
\DpName{P.Adzic}{DEMOKRITOS}
\DpName{Z.Albrecht}{KARLSRUHE}
\DpName{T.Alderweireld}{AIM}
\DpName{G.D.Alekseev}{JINR}
\DpName{R.Alemany}{VALENCIA}
\DpName{T.Allmendinger}{KARLSRUHE}
\DpName{P.P.Allport}{LIVERPOOL}
% ...
\normalsize
\endgroup
\titlefoot{Department of Physics and Astronomy, Iowa State
     University, Ames IA 50011-3160, USA
    \label{AMES}}
\titlefoot{Physics Department, Univ. Instelling Antwerpen,
     Universiteitsplein 1, B-2610 Antwerpen, Belgium \\
     \indent~~and IIHE, ULB-VUB,
     Pleinlaan 2, B-1050 Brussels, Belgium \\
     \indent~~and Facult\'e des Sciences,
     Univ. de l'Etat Mons, Av. Maistriau 19, B-7000 Mons, Belgium
    \label{AIM}}
\titlefoot{Physics Laboratory, University of Athens, Solonos Str.
     104, GR-10680 Athens, Greece
    \label{ATHENS}}
\titlefoot{Department of Physics, University of Bergen,
     All\'egaten 55, NO-5007 Bergen, Norway
    \label{BERGEN}}
\titlefoot{Dipartimento di Fisica, Universit\`a di Bologna and INFN,
     Via Irnerio 46, IT-40126 Bologna, Italy
    \label{BOLOGNA}}
\titlefoot{Centro Brasileiro de Pesquisas F\'{\i}sicas, rua Xavier Sigaud 150,
     BR-22290 Rio de Janeiro, Brazil \\
     \indent~~and Depto. de F\'{\i}sica, Pont. Univ. Cat\'olica,
     C.P. 38071 BR-22453 Rio de Janeiro, Brazil \\
     \indent~~and Inst. de F\'{\i}sica, Univ. Estadual do Rio de Janeiro,
     rua S\~{a}o Francisco Xavier 524, Rio de Janeiro, Brazil
    \label{BRASIL}}
\titlefoot{Comenius University, Faculty of Mathematics and Physics,
     Mlynska Dolina, SK-84215 Bratislava, Slovakia
    \label{BRATISLAVA}}
\titlefoot{Coll\`ege de France, Lab. de Physique Corpusculaire, IN2P3-CNRS,
     FR-75231 Paris Cedex 05, France
    \label{CDF}}
\titlefoot{CERN, CH-1211 Geneva 23, Switzerland
    \label{CERN}}
\titlefoot{Institut de Recherches Subatomiques, IN2P3 - CNRS/ULP - BP20,
     FR-67037 Strasbourg Cedex, France
    \label{CRN}}
\titlefoot{Now at DESY-Zeuthen, Platanenallee 6, D-15735 Zeuthen, Germany
    \label{DESY}}
\titlefoot{Institute of Nuclear Physics, N.C.S.R. Demokritos,
     P.O. Box 60228, GR-15310 Athens, Greece
    \label{DEMOKRITOS}}
\titlefoot{FZU, Inst. of Phys. of the C.A.S. High Energy Physics Division,
     Na Slovance 2, CZ-180 40, Praha 8, Czech Republic
    \label{FZU}}
\titlefoot{Dipartimento di Fisica, Universit\`a di Genova and INFN,
     Via Dodecaneso 33, IT-16146 Genova, Italy
    \label{GENOVA}}
\titlefoot{Institut des Sciences Nucl\'eaires, IN2P3-CNRS, Universit\'e
     de Grenoble 1, FR-38026 Grenoble Cedex, France
    \label{GRENOBLE}}
\titlefoot{Helsinki Institute of Physics, HIP,
     P.O. Box 9, FI-00014 Helsinki, Finland
    \label{HELSINKI}}
\titlefoot{Joint Institute for Nuclear Research, Dubna, Head Post
     Office, P.O. Box 79, RU-101 000 Moscow, Russian Federation
    \label{JINR}}
\titlefoot{Institut f\"ur Experimentelle Kernphysik,
     Universit\"at Karlsruhe, Postfach 6980, DE-76128 Karlsruhe,
     Germany
    \label{KARLSRUHE}}
\titlefoot{Institute of Nuclear Physics and University of Mining and Metalurgy,
     Ul. Kawiory 26a, PL-30055 Krakow, Poland
    \label{KRAKOW}}
\titlefoot{Universit\'e de Paris-Sud, Lab. de l'Acc\'el\'erateur
     Lin\'eaire, IN2P3-CNRS, B\^{a}t. 200, FR-91405 Orsay Cedex, France
    \label{LAL}}
\titlefoot{School of Physics and Chemistry, University of Lancaster,
     Lancaster LA1 4YB, UK
    \label{LANCASTER}}
\titlefoot{LIP, IST, FCUL - Av. Elias Garcia, 14-$1^{o}$,
     PT-1000 Lisboa Codex, Portugal
    \label{LIP}}
\titlefoot{Department of Physics, University of Liverpool, P.O.
     Box 147, Liverpool L69 3BX, UK
    \label{LIVERPOOL}}
\titlefoot{LPNHE, IN2P3-CNRS, Univ.~Paris VI et VII, Tour 33 (RdC),
     4 place Jussieu, FR-75252 Paris Cedex 05, France
    \label{LPNHE}}
\titlefoot{Department of Physics, University of Lund,
     S\"olvegatan 14, SE-223 63 Lund, Sweden
    \label{LUND}}
\titlefoot{Universit\'e Claude Bernard de Lyon, IPNL, IN2P3-CNRS,
     FR-69622 Villeurbanne Cedex, France
    \label{LYON}}
\titlefoot{Univ. d'Aix - Marseille II - CPP, IN2P3-CNRS,
     FR-13288 Marseille Cedex 09, France
    \label{MARSEILLE}}
\titlefoot{Dipartimento di Fisica, Universit\`a di Milano and INFN-MILANO,
     Via Celoria 16, IT-20133 Milan, Italy
    \label{MILANO}}
\titlefoot{Dipartimento di Fisica, Univ. di Milano-Bicocca and
     INFN-MILANO, Piazza delle Scienze 2, IT-20126 Milan, Italy
    \label{MILANO2}}
\titlefoot{Niels Bohr Institute, Blegdamsvej 17,
     DK-2100 Copenhagen {\O}, Denmark
    \label{NBI}}
\titlefoot{IPNP of MFF, Charles Univ., Areal MFF,
     V Holesovickach 2, CZ-180 00, Praha 8, Czech Republic
    \label{NC}}
\titlefoot{NIKHEF, Postbus 41882, NL-1009 DB
     Amsterdam, The Netherlands
    \label{NIKHEF}}
\titlefoot{National Technical University, Physics Department,
     Zografou Campus, GR-15773 Athens, Greece
    \label{NTU-ATHENS}}
\titlefoot{Physics Department, University of Oslo, Blindern,
     NO-1000 Oslo 3, Norway
    \label{OSLO}}
\titlefoot{Dpto. Fisica, Univ. Oviedo, Avda. Calvo Sotelo
     s/n, ES-33007 Oviedo, Spain
    \label{OVIEDO}}
\titlefoot{Department of Physics, University of Oxford,
     Keble Road, Oxford OX1 3RH, UK
    \label{OXFORD}}
\titlefoot{Dipartimento di Fisica, Universit\`a di Padova and
     INFN, Via Marzolo 8, IT-35131 Padua, Italy
    \label{PADOVA}}
\titlefoot{Rutherford Appleton Laboratory, Chilton, Didcot
     OX11 OQX, UK
    \label{RAL}}
\titlefoot{Dipartimento di Fisica, Universit\`a di Roma II and
     INFN, Tor Vergata, IT-00173 Rome, Italy
    \label{ROMA2}}
\titlefoot{Dipartimento di Fisica, Universit\`a di Roma III and
     INFN, Via della Vasca Navale 84, IT-00146 Rome, Italy
    \label{ROMA3}}
\titlefoot{DAPNIA/Service de Physique des Particules,
     CEA-Saclay, FR-91191 Gif-sur-Yvette Cedex, France
    \label{SACLAY}}
\titlefoot{Instituto de Fisica de Cantabria (CSIC-UC), Avda.
     los Castros s/n, ES-39006 Santander, Spain
    \label{SANTANDER}}
\titlefoot{Dipartimento di Fisica, Universit\`a degli Studi di Roma
     La Sapienza, Piazzale Aldo Moro 2, IT-00185 Rome, Italy
    \label{SAPIENZA}}
\titlefoot{Inst. for High Energy Physics, Serpukov
     P.O. Box 35, Protvino, (Moscow Region), Russian Federation
    \label{SERPUKHOV}}
\titlefoot{J. Stefan Institute, Jamova 39, SI-1000 Ljubljana, Slovenia
     and Laboratory for Astroparticle Physics,\\
     \indent~~Nova Gorica Polytechnic, Kostanjeviska 16a, SI-5000 Nova Gorica, Slovenia, \\
     \indent~~and Department of Physics, University of Ljubljana,
     SI-1000 Ljubljana, Slovenia
    \label{SLOVENIJA}}
\titlefoot{Fysikum, Stockholm University,
     Box 6730, SE-113 85 Stockholm, Sweden
    \label{STOCKHOLM}}
\titlefoot{Dipartimento di Fisica Sperimentale, Universit\`a di
     Torino and INFN, Via P. Giuria 1, IT-10125 Turin, Italy
    \label{TORINO}}
\titlefoot{Dipartimento di Fisica, Universit\`a di Trieste and
     INFN, Via A. Valerio 2, IT-34127 Trieste, Italy \\
     \indent~~and Istituto di Fisica, Universit\`a di Udine,
     IT-33100 Udine, Italy
    \label{TU}}
\titlefoot{Univ. Federal do Rio de Janeiro, C.P. 68528
     Cidade Univ., Ilha do Fund\~ao
     BR-21945-970 Rio de Janeiro, Brazil
    \label{UFRJ}}
\titlefoot{Department of Radiation Sciences, University of
     Uppsala, P.O. Box 535, SE-751 21 Uppsala, Sweden
    \label{UPPSALA}}
\titlefoot{IFIC, Valencia-CSIC, and D.F.A.M.N., U. de Valencia,
     Avda. Dr. Moliner 50, ES-46100 Burjassot (Valencia), Spain
    \label{VALENCIA}}
\titlefoot{Institut f\"ur Hochenergiephysik, \"Osterr. Akad.
     d. Wissensch., Nikolsdorfergasse 18, AT-1050 Vienna, Austria
    \label{VIENNA}}
\titlefoot{Inst. Nuclear Studies and University of Warsaw, Ul.
     Hoza 69, PL-00681 Warsaw, Poland
    \label{WARSZAWA}}
\titlefoot{Fachbereich Physik, University of Wuppertal, Postfach
     100 127, DE-42097 Wuppertal, Germany
    \label{WUPPERTAL}}
\addtolength{\textheight}{-10mm}
\addtolength{\footskip}{5mm}
\clearpage
\headsep 30.0pt
\end{titlepage}
%
\pagenumbering{arabic} 
\setcounter{footnote}{0} %
\large
%
%
%
\newcommand{\EM}{electromagnetic }
\newcommand{\bb}{$b \bar{b}\ $}
\newcommand{\Lb}{$\Lambda_b\ $}
\newcommand{\Sb}{$\Sigma_b\ $}
\newcommand{\Xb}{$\Xi_b\ $}
\newcommand{\Ob}{$\Omega_b\ $}
\newcommand{\Lc}{$\Lambda_c $}
\newcommand{\Sc}{$\Sigma_c\ $}
\newcommand{\Xc}{$\Xi_c\ $}
\newcommand{\Oc}{$\Omega_c\ $}
\newcommand{\Ls}{$\Lambda$}
\newcommand{\Ll}{$\Lambda \ell\ $}
\newcommand{\Lcl}{$\Lambda_c \ell\ $}
\newcommand{\Z}{$Z^0\ $}
\newcommand{\Vcb}{\ifmmode {|V_{cb}|}\else {$|V_{cb}|$}\fi}
\newcommand{\Vub}{\ifmmode {|V_{ub}|}\else {$|V_{ub}|$}\fi}
\newcommand{\om} {\ifmmode w \else {\it w} \fi}
\newcommand{\CKM}{Cabibbo-Kobayashi-Maskawa}
\newcommand{\mbb}{b \bar{b}}
\newcommand{\mLb}{\Lambda_b}
\newcommand{\mSb}{\Sigma_b}
\newcommand{\mXb}{\Xi_b}
\newcommand{\mOb}{\Omega_b}
\newcommand{\mLc}{\Lambda_c}
\newcommand{\mSc}{\Sigma_c}
\newcommand{\mXc}{\Xi_c}
\newcommand{\mOc}{\Omega_c}
\newcommand{\mLs}{\Lambda}
\newcommand{\mLl}{\Lambda \ell}
\newcommand{\mLcl}{\Lambda_c \ell}
\newcommand{\mZ}{Z^0}
\newcommand{\ra}{\rightarrow} 
\newcommand{\bpro}{$b-proton$}
\newcommand{\bapro}{$b-antiproton$}
\newcommand{\blp}{\mbox{$\rm b\rightarrow\nb\rightarrow l^-p^+X$}}
\newcommand{\zqq}[1]{\mbox{ $\rm Z^0\rightarrow {#1}\bar{#1}$} }
\newcommand{\Lam}[1]{\mbox{$\rm \Lambda_{#1}$} }
\newcommand{\nb}{\mbox{$\rm N_b$} }
\newcommand{\ptr}{\mbox{ $p_t$} }
\newcommand{\dv}{\mbox{$\delta_V$}}
\newcommand{\mup}{\mbox{$\mu$\rm p}}
\newcommand{\muK}{\mbox{$\mu$\rm K}}
\newcommand{\mupi}{\mbox{$\mu\pi$}}
\newcommand{\muX}{\mbox{$\mu$\rm X}}
\newcommand{\GeVc}{\mbox{\rm G$e$V/$c$}}
\newcommand{\DEDX}{\mbox{${dE/dx}$}}
\newcommand{\dedx}{\mbox{$\frac{dE}{dx}$}}
\newcommand{\sion}{\mbox{$\sigma_{\frac{dE}{dx} }$}}
\newcommand{\idr}{\mbox{\it id$_{\rm RICH}$}}
\newcommand{\itt}{\mbox{$\diamondsuit$}}
\newcommand{\isdef}{\mbox{--}}
\newcommand{\sgn}{\mbox{\it sign}}
\newcommand{\selm}{\mbox{$\cal M$}}
\newcommand{\selh}{\mbox{$\cal H$}}
\newcommand{\selev}{\mbox{$\cal E$}}
\newcommand{\selv}{\mbox{$\cal V$}}
\newcommand{\signal}{\mbox{$\mu$p from $b$-baryon}}
\newcommand{\thetc}{\mbox{$\rm\vartheta_C$}}
\newcommand{\ptm}{\mbox{$ p_{T}$}}
\newcommand{\pts}{\mbox{$ p_{T}^{(S)}$}}
\newcommand{\tplb}{\mbox{$\tau_{\mu\mbox{\scriptsize p}}$}}
\newcommand{\tbck}{\mbox{$\tau_{\mbox{\scriptsize BGD}}$}}
\newcommand{\fbck}{\mbox{$f_{\mbox{\scriptsize BGD}}$}}
\newcommand{\qb}{\mbox{b}}
\newcommand{\Bb}{\mbox{$b$-baryon}}
\newcommand{\Br}{\ifmmode {\mathrm{BR}} \else {\mbox{BR}} \fi} 
\newcommand{\munu}{\mbox{$\mu\bar{\nu}_{\mu}$}}
\newcommand{\stat}{\mbox{(stat.)}}
\newcommand{\syst}{\mbox{(syst.)}}
\newcommand{\prot}{\mbox{p}}
\newcommand{\X}{\mbox{X}}
\newcommand{\corrhyp}{\mbox{${C}_{\Lambda}$}}
\newcommand{\corrM}{\mbox{${C}_{\cal M}$}}
\newcommand{\effsel}{\mbox{$\epsilon_{ {\cal H}\wedge{\cal V} } $ }}
\newcommand{\classes}{\mbox{\it classes}}
\newcommand{\class}{\mbox{\it class}}
\newcommand{\samples}{\mbox{\it samples}}
\newcommand{\sample}{\mbox{\it sample}}
\newcommand{\event}{\mbox{\it event}}
\newcommand{\sigmat}{\mbox{$\sigma_t$}}
\newcommand{\Pev}{\mbox{${\cal P}_{\event}$}}
\newcommand{\Pevi}{\mbox{${\cal P}(\event|i)$}}
\newcommand{\Pdx}{\mbox{${\cal P}_{\frac{dE}{dx}}$}}
\newcommand{\Ppt}{\mbox{${\cal P}_{\perp}$}}
\newcommand{\Pti}{\mbox{${\cal P}_{t}$}}
\newcommand{\Pfr}{\mbox{${\cal F}$}}
\newcommand{\Aone}{\mbox{${\cal A}_1$}}
\newcommand{\rha}{\mbox{$\rho^2_{{\cal A}_1}$}}
\newcommand{\rhf} {\mbox{$\rho^2_{\cal F}$}}
\newcommand{\Sgn}{\mbox{${\cal S_L}$}}
\newcommand{\Rss}{\mbox{${\cal R}^{**}$}}
\newcommand{\Nss}{\mbox{${\cal N}^{**}$}}
\newcommand{\Ns}{\mbox{${\cal N}^{*}$}}
\newcommand{\Rj}{\mbox{${\cal R}_{f}$}}
\newcommand{\pdf}{\mbox{${\it pdf}$}}
\newcommand{\bt}{\begin{table}}
\newcommand{\et}{\end{table}}
\newcommand{\bi}{\begin{itemize}}
\newcommand{\ei}{\end{itemize}}
\newcommand{\bc}{\begin{center}}
\newcommand{\ec}{\end{center}}
\newcommand{\be}{\begin{equation}}
\newcommand{\ee}{\end{equation}}
\newcommand{\ba}{\begin{eqnarray}}
\newcommand{\ea}{\end{eqnarray}}
\newcommand{\Bz}{\ifmmode {\bar{B_d^0}} \else {\mbox{$\bar{B_d^0}$}} \fi}
\newcommand{\ptl}{\mbox{$p_t^\ell$}}
\newcommand{\Ds}{\ifmmode {D^{*+}} \else {\mbox{$D^{*+}$}} \fi}
\newcommand{\dm}{\mbox{$\Delta m$}}
\newcommand{\D} {\ifmmode{\Delta} \else {$\Delta$} \fi}
\newcommand{\Dz}{\mbox{$D^{0}$}}
\newcommand{\ps}{\mbox{$\pi^{*}$}}
\newcommand{\Dss}{\mbox{$D^{**}$}}
\newcommand{\mm}{\mbox{$\mu^2$}}
\newcommand{\BtoDs}{\mbox{$\bar{B^0_d}\rightarrow D^{*+} \ell^- \bar{\nu_\ell}$}}
\newcommand{\BtoDss}{\mbox{$b\rightarrow D^{**} \ell \bar{\nu_\ell}$}}

%
%
\section{Introduction}

In the framework of the Standard Model, the mixing between quarks of different flavours
is described by the Cabibbo-Kobayashi-Maskawa (CKM) matrix. 
Its elements are not predicted 
by the theory, apart from the constraints due to the requirement of unitarity.\par
A precise measurement of \Vcb, the element corresponding to the beauty to charm quark transitions, 
constrains the parameters which describe the
process of CP violation  for \Bz mesons\footnote{Charge conjugated states are always implied; 
lepton ($\ell$)
means either an electron or a muon, unless the contrary is explicitly stated.} 
\cite{Achille}. 
As a result of progress in the phenomenological description of heavy flavour 
semileptonic decays, \Vcb\ is determined with small theoretical uncertainty, 
from either the inclusive process
$b \rightarrow c\ell^- \bar\nu_{\ell}$, or from an analysis of the form factors in the decay
\BtoDs . The present measurement is based on the second approach \cite{method}.\par
The decay rate for the last process is proportional to \Vcb$^2$ and to the hadron matrix elements
describing the transition from a \Bz to a \Ds meson.
In the limit of very heavy quarks ($m_{b,c}~ >> ~ \Lambda_{QCD}~ \sim~ 200~ {\mathrm{MeV}}/c^2$), 
the amplitude is proportional to a single form factor \Pfr(\om), where \om 
is the scalar product of the \Bz and \Ds four-velocities. It is equal to the \Ds 
Lorentz $\gamma$ factor  in the \Bz rest frame.
When \mbox{\om = 1}, the \Ds is produced at rest in the \Bz rest frame:
as a consequence of Heavy Flavour symmetry, the normalisation \Pfr(1) = 1 is expected.
Corrections to this prediction due to perturbative QCD have been computed up to second order \cite{QCDII}.
The effect of finite $b,c$ quark masses has been calculated in the framework of the Heavy Quark 
Effective Theory \cite{F1}. 
The value \Pfr(1) = 0.91 $\pm$ 0.03, as determined in \cite{Neuold}, was used in this analysis.
This result is consistent with the value \Pfr(1) = 0.88 $\pm$ 0.05, derived in \cite{LEPBWG} 
on the basis of the study presented in \cite{Bigi}, and with a more recent computation
based on lattice QCD which yielded \Pfr(1) = $0.93 \pm 0.03$~\cite{Simone}.
\par
The measurement of the decay rate at \om = 1 would therefore determine \Vcb\ with
small theoretical uncertainty. Due to phase space suppression, this quantity is determined from
the extrapolation to 1 of the differential decay rate d$\Gamma$/d\om, 
where \Pfr(\om) is parametrised according to several different functional forms \cite{Neuold,BGL,Neunew} 
(see also discussion below). 
Results based on this approach have been reported by the ARGUS \cite{ARGUS} and CLEO \cite{CLEO,CLEO2}
collaborations operating at the $\Upsilon$(4S) resonance, and by ALEPH \cite{ALEPH1,ALEPH2} and 
OPAL \cite{OPAL,OPAL2} at LEP. 
The present paper updates the previous DELPHI result of reference \cite{DELPHI:vcb}. Identification
of \Ds mesons is based on the tagging of the soft pion (\ps) from the decay 
\mbox{\Ds $\rightarrow~ D^0 \pi^+~$}, the method referred to as ``inclusive analysis'' 
in reference \cite{DELPHI:vcb}. 
As compared to this previous work, the following improvements were obtained:
\bi 
\item the resolution on \om was improved by a factor of about 1.5 by applying the 
algorithm of inclusive secondary vertex reconstruction developed for \Bz lifetime 
\cite{DELPHI:taubo} and oscillation \cite{DELPHI:bosc} measurements;
\item the full available statistics was analysed, thereby increasing the sample by
more than a factor two;
\item the most recent parametrisation \cite{Neunew} of \Pfr(\om) was used to extrapolate the
experimental data to \om=1;
\item a more precise determination of the \mbox{BR($b\ra\ell^- \bar{\nu_{\ell}} \Ds X$)} was used to 
compute the fraction of events in the sample due to 
non-resonant \Ds$\pi$ production, or to the intermediate production
of higher excited charm states which then decay into a \mbox{\Ds;}
all these states will be called \Dss\ in the following. 
\ei

\section{The DELPHI detector }

The DELPHI detector has been described in detail elsewhere 
\cite{DELPHI:dete}.
Charged particle tracking through the uniform axial magnetic field (B = 1.23 T), 
secondary vertex reconstruction and lepton identification are important in
this analysis: they will be briefly described in the following. \par
The detector elements used for tracking are the Vertex
Detector (VD), the Inner Detector (ID), the Time Projection Chamber
(TPC), the Outer Detector (OD) in the barrel and the Forward Chambers in 
the endcap regions. 
The average momentum resolution for high momentum ($p$) charged particles 
in the polar angle region between 30$^\circ$ and
150$^\circ$ is $\sigma(p)/p = 0.0006 ~p~ $ (GeV/$c$) \cite{DELPHI:dete}. \par
The VD, consisting of 3 cylindrical layers of silicon detectors
(radii 6, 8 and 11 cm), provides up to 3 hits per track
(or more in small overlapping regions) in the polar angle range
$ 43^\circ < \theta < 137^\circ$. In the original design the VD provided only
two-dimensional information in the $R\phi$ plane, orthogonal to the beam
direction. Since the 1994 data taking, an upgraded detector with full three-dimensional
 point reconstruction was installed. In the $R\phi$ plane the spatial resolution
of the VD is about 8~$\mu$m per point. Tracks from charged particles 
are extrapolated back to the beam collision point with a resolution of 
\mbox{$\sqrt{20^2+65^2/p^2_\perp}$}~$\mu$m, where $p_\perp$ is the momentum of the
particle in the $R\phi$ plane. The resolution on the {\it z} coordinate depends on {\it z}
and is on average slightly worse than that in $R\phi$. 
 The primary vertex of the $e^+e^-$ interaction was reconstructed on an
 event-by-event basis using a  beam spot constraint. The position of the primary vertex
 could be determined in this way with an average precision of about 40~$\mu$m
 (slightly dependent on the flavour of the primary quark-antiquark pair) in 
 the plane transverse to the beam direction. 
 Secondary vertices from B semileptonic decays were reconstructed with 
 high efficiency employing the algorithm described in reference \cite{DELPHI:taubo}.
 The decay length resolution for the present analysis was about 400~$\mu$m. \par
 Leptons were identified among all the charged particles of momentum $2 < p < 30~$ GeV/$c$.
 To allow the reconstruction of the \Bz decay point only particles with at least one hit in the 
 VD were considered as lepton candidates.\par
 Electron identification was based on a neural network algorithm, optimally combining
 the information from the ionisation signal in the TPC, from the energy
 release in the electromagnetic calorimeters, and, for tracks with momentum
 below 3 GeV/$c$, from the Ring Imaging CHerenkov counters (RICH). A level of tagging 
 providing about $75 \%$ efficiency within the calorimeter acceptance was chosen. 
 The probability for a hadron to fake an electron was about $1 \%$.
 Electrons from photon conversions are mainly produced in the outer ID wall and in the inner 
 TPC frame. About 80\% of them were removed with negligible loss of signal by 
 reconstructing their materialisation vertex.\par
 Muons were selected by matching the track reconstructed in the tracking system to the 
 track elements provided by the barrel and forward muon chambers. The efficiency was about
 80\% for about 1\% probability of hadron mis-identification.\par
 The experimental efficiencies and hadron mis-identification  probabilities were 
 measured year by year  using dedicated samples of leptons and hadrons 
 independently tagged
 and the simulation was tuned consequently.

\section{Hadronic Event Selection and Simulation}

 Charged particles were required to have a momentum in the range \mbox{
 $0.25 < p < 45$~ GeV/$c$}, a relative error on the momentum measurement less than \mbox{100\%},
 a distance of closest approach to the interaction point less than 10 cm in 
$R\phi$ and 25 cm along {\it z}, and
 a polar angle such that $\mid$cos$\theta\mid <$~0.937.
 Electromagnetic showers not associated to tracks were required to be well contained
 within the calorimeter acceptance and to have an energy release
 greater than 0.5 (0.3) GeV in the barrel (forward) electromagnetic calorimeter.
 Only hadronic showers with an energy release greater than 1 GeV and not associated to tracks from
 charged particles were accepted as neutral hadrons.\par 
 The following selection was applied to the detector operating conditions: the TPC was required to be 
 fully efficient,  and at least \mbox{95\%} of the 
 electromagnetic calorimeters and 90\% of the muon chambers had to be active.
 Hadronic $Z$ decays were selected with 95\% efficiency and negligible background by using 
 standard cuts (see reference \cite{DELPHI:dete}).  \par
 Each event was divided into two opposite hemispheres by a plane 
 orthogonal to the thrust axis.  To ensure that the event was well contained 
 inside the fiducial volume of the detector the polar angle of the thrust axis of the event
 had to satisfy the requirement $\mid$cos$\theta\mid<$  0.95.
 Charged and neutral particles were clustered into jets by using the
 LUCLUS \cite{JETSET} algorithm with default resolution parameter {\it djoin } = 2.5 GeV/$c$. \par
 About three million events were selected from the full LEP I data sets. The JETSET 7.3 
 Parton Shower \cite{JETSET} program was used to generate hadronic $Z$ decays, which were followed
 through the detailed detector simulation DELSIM \cite{DELSIM} and finally processed by the 
 same analysis chain as the real data. A sample of about seven million $Z$ $\rightarrow q\bar{q}$ events
 was used. To increase the statistical significance of the simulation, an additional sample of
 about 2.2 million $Z$ $\rightarrow b\bar{b}$ was analysed, equivalent to about ten million 
 hadronic $Z$ decays. Details of the $Z$ samples used are given in Table \ref{tab:qqstat}.

\bt[htb]
\bc
\begin{tabular} {|c|c|c|c|}
\hline
 Year         & real data  & simulated  & simulated \\
              &            & $Z\rightarrow~q\bar{q}$ &  $Z\rightarrow~b\bar{b}$ \\  \hline
 1992+1993    & ~1203982~  & ~2012615~               &~~922764~ \\
 1994+1995    & ~1832082~  & ~5190586~               &~1321384~ \\
 Total        & ~3036064~  & ~7203201~               &~2244148~ \\ \hline
\end{tabular}
\caption[]{Available number of events. In 1992 and 1993 only two-dimensional vertex 
 reconstruction was available.}
\label{tab:qqstat}
\ec
\et

\boldmath
\section{The \Ds $\ell^-$ $\bar{\nu}$ sample}
\unboldmath

\subsection{Event Selection} 
\label{sec:selection}
 Only events containing at least one lepton candidate were considered further. 
 The transverse momentum of the lepton  relative to the jet it belonged to, \ptl, was computed after
 removing the lepton from the jet. The cut \ptl~ $>$ 1 GeV/$c$ was imposed to 
reduce the background.\par
 A charm hadron candidate was reconstructed from all the particles in the jet containing the lepton,
 except the lepton itself, by means of the iterative algorithm described in detail in 
reference \cite{DELPHI:taubo}.
 Small clusters were first formed out of the charged particles and, 
 when possible, a decay vertex was computed for each cluster. The charm candidate so obtained
 was intersected with the lepton trajectory to provide the \Bz secondary vertex. 
 In the case where only one charged particle with hits in the VD belonged to 
 the cluster, its intersection with the lepton track was computed. 
 The cluster associated to the secondary vertex with the largest statistical significance \Sgn\
 (defined as the distance from the primary vertex divided by its error; in years 1992 and 1993 
 only the projected distance onto the $R\phi$ plane was considered) was kept as a seed.
 All other charged and neutral particles in the jet were ordered by decreasing values of their
 pseudo-rapidity relative to the cluster direction, and added to it 
 provided the mass of the system did not exceed 2.2 GeV/$c^2$. The charm three-momentum
 was obtained from the sum of all the particles assigned to the cluster.
 The charm trajectory was evaluated again and was finally intersected with the 
 lepton track to obtain the \Bz decay point. To improve background rejection and the resolution
 on \om (see below), events with significance \Sgn $<$4.5 were rejected. \par
\begin{figure}[b]
\begin {center}
\setlength{\unitlength}{1cm}
\begin{picture}(15,15)
\epsfxsize=15cm
\epsffile{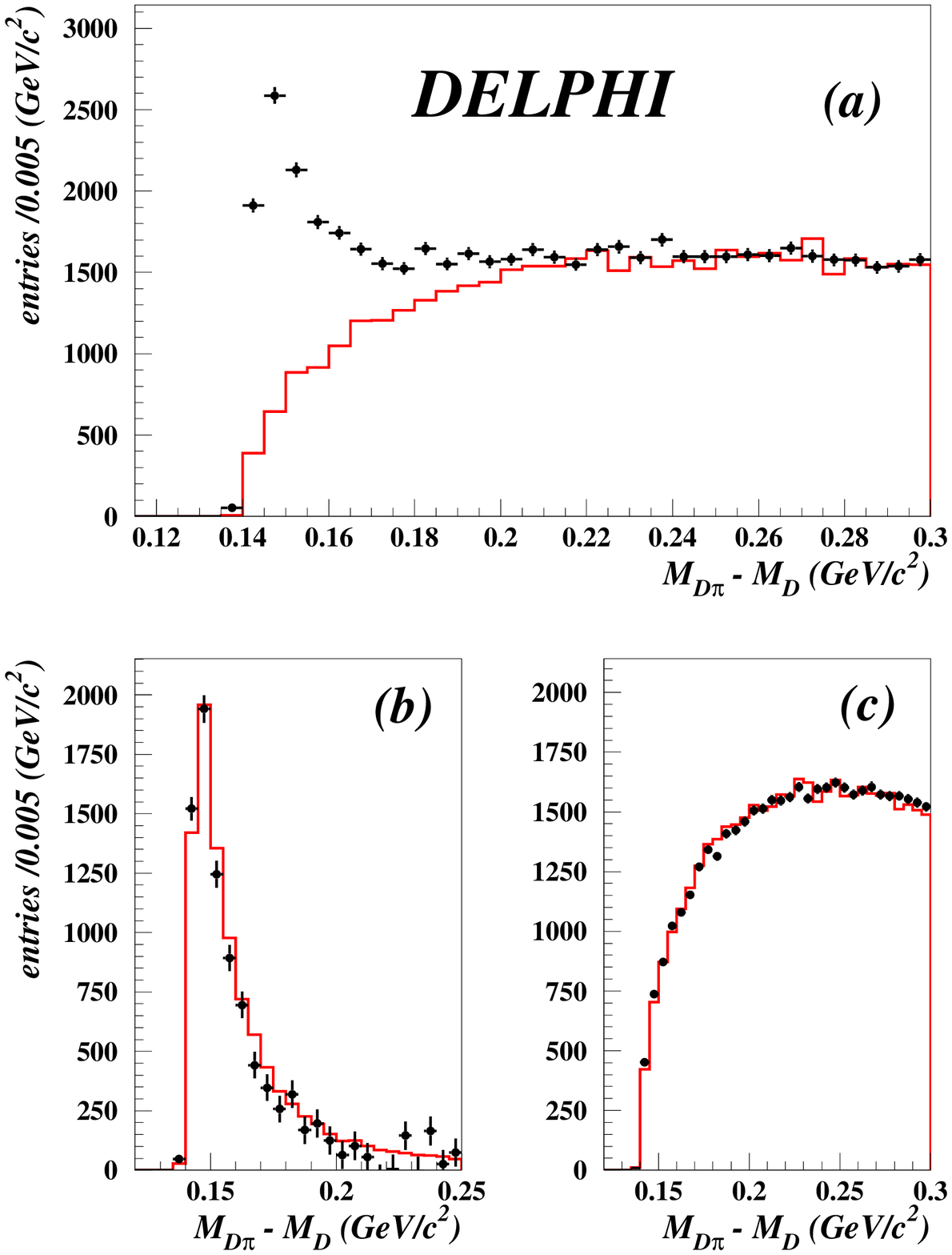}
\end{picture}
\end{center}
\caption[]{ Mass difference $M_{D\pi}-M_{D}$. \\
a) opposite charge, real data (dots with error bars), 
same charge (shaded area), normalised to the side band defined in the text. The \Ds\ 
signal is clearly visible.\\
b) opposite charge, real data after subtraction of the combinatorial 
background (dots with error bars). This agrees well with the resonant 
contribution from simulation (shaded area); \\
c) simulation: combinatorial background from opposite charge 
(dots),  which is consistent with the same charge 
combinations normalised in the side band (shaded area).}

\label{fig:deltam}
\end{figure}

 The \ps\ candidate (the pion from \Ds decay) was searched for among all 
 particles in the jet with charge opposite 
 to that of the lepton. If the candidate belonged to the charm cluster, the 
 \Dz~ four-momentum
 was computed after removing the \ps\ from the cluster and imposing the \Dz~ mass.
 To increase efficiency, particles classified as fragmentation products were 
 also considered as
 \ps\ candidates. The \Dz~ was then identified with the charm cluster, constrained to the \Dz~ mass. 
 \Ds production was finally tagged based on the mass difference \mbox{\dm = M$_{\Dz\ps}$ - M$_{\Dz}$} 
 (see Figure \ref{fig:deltam}).
 All events with \dm $<$0.165 GeV/$c^2$ were used for the analysis. 

\subsection{Event Kinematics}
\label{sec:kine}

 The variable \om (= $v_\Bz \cdot v_\Ds$) can be expressed as:
\bc  \om = $(M_{\Bz}^2 + M_{\Ds}^2 - q^2)/(2M_{\Bz} M_{\Ds}) $ \ec
 where $q^2$ is obtained from the \Bz and \Ds four-momenta as: 
\bc  \[q^2 = (p_{\Bz}-p_{\Ds})^2  \]\ec
 The \Ds energy, polar and azimuthal angles, and the energy of the \Bz meson were determined as in 
reference \cite{DELPHI:vcb}. The resolution obtained in the simulation was:
\ba
\nonumber &\sigma(E_{\Bz}) / E_{\Bz} &= ~10\% \\
\nonumber &\sigma(E_{\Ds}) / E_{\Ds} &= ~12\% \\
\nonumber &\sigma(\theta_{\Ds})      &= ~18 ~\mathrm{mrad}\\
\nonumber &\sigma(\phi_{\Ds})        &= ~21 ~\mathrm{mrad}
\ea
\par The \Bz direction was evaluated using two estimators:
\bi
\item the direction obtained by inverting the vector sum of all the particles in the event except the ones
assigned to the \Bz. This procedure, already used in reference \cite{DELPHI:vcb}, exploits three-momentum conservation 
 in the event. The resolution depends on the hermeticity of the detector, but can also be spoiled whenever 
 another semileptonic decay takes place in the event;
\item the direction of the vector joining the primary and the secondary vertex: the resolution achieved
 depends on the distance between the two vertices, improving for higher values. This approach was not used in
 the inclusive analysis of reference \cite{DELPHI:vcb}.
\ei
Using the simulation, the resolution was parametrised on the basis of the missing energy in the  
hemisphere opposite to the \Bz for the first estimator, and as a function of the reconstructed 
decay distance of the \Bz for the second. The \Bz meson direction was then obtained as the average of the
two, weighted by the inverse of their error squared. When the difference between the two values  
was greater than three times its error, the direction nearer to the \Ds $\ell^-$ system was chosen.
In the years 1992 and 1993 only the first estimator could be used to determine the \Bz polar angle.
The resolution function obtained could be parametrised by a Breit-Wigner distribution, with 
half width at half maximum:
\ba
\nonumber &\Gamma(\phi_{\Bz})/2   &= ~12 ~\mathrm{mrad}\\
\nonumber &\Gamma(\theta_{\Bz})/2 &= ~12 ~\mathrm{mrad}  ~~(1994-1995)\\
\nonumber &\Gamma(\theta_{\Bz})/2 &= ~24 ~\mathrm{mrad}  ~~(1992-1993)
\ea
\par
\begin{figure}[b]
\begin {center}
\setlength{\unitlength}{1cm}
\begin{picture}(15,15)
\epsfxsize=15cm
\epsffile{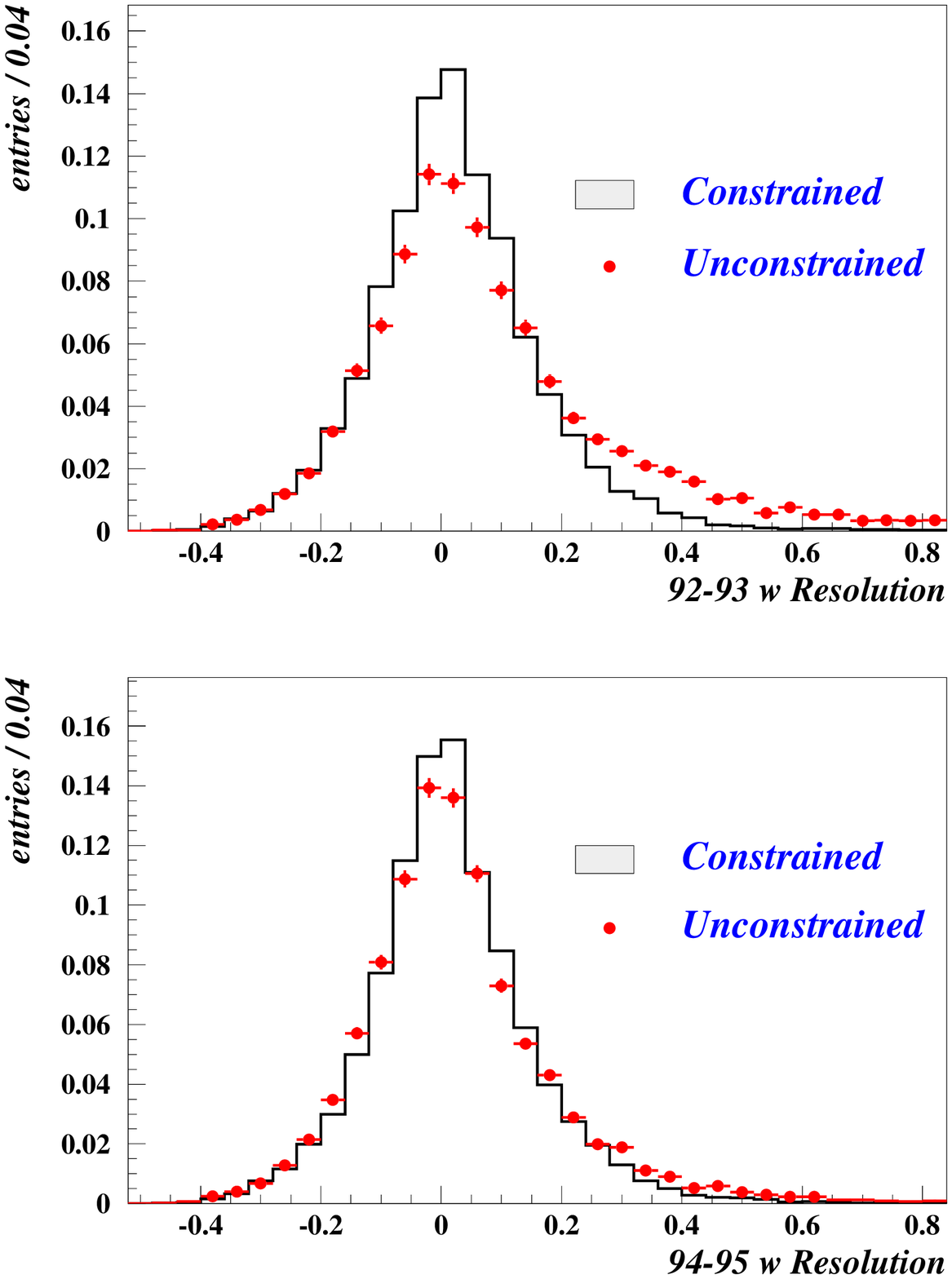}
\end{picture}
\end{center}
\caption[]{ $\om$ resolution. Upper plot : 1992-1993 analysis; lower plot: 
1994-1995 analysis. \\
Dots: experimental resolution without exploiting kinematic 
constraints. Since 1994 three-dimensional vertex reconstruction helped 
improve the resolution. \\ 
Shaded area: further improvement due to the requirement \mm = 0 (see text).}

\label{fig:omgres}
\end{figure}

The resulting \om resolution function is shown by the dots in Figure \ref{fig:omgres}. 
The RMS width of the core of the
distribution is approximately the same for all data sets ($\sigma$(\om) = 
0.125), but larger tails
are present in the 1992-1993 sample due to the poorer $\theta$ measurement. The RMS width corresponds to 
about 25\% of the allowed kinematic range (1 $<$ \om ~$<$~ 1.504). Due to resolution effects 
(17.9 $\pm$ 0.4)\% ((32.9$\pm$0.6)\%) of the events of the 94-95 (92-93) data 
set lay outside that range. \par
The squared recoil mass \mm~ was also determined on the basis of the event kinematics. It is 
defined as:
\be
\mm = M_{\Bz}^2 +M^2_{\Ds \ell^-}-2 P_{\Bz} \cdot P_{\Ds \ell^-}, 
\label{eq:mm2}
\ee
where $M_{\Bz(\Ds \ell^-)},P_{\Bz(\Ds \ell^-)}$ are the mass and four momenta of the \Bz meson
and \mbox{\Ds$\ell^-$}~ system respectively. In the decay process \BtoDs, \mm~ represents the square 
of the mass of the neutrino, and should be zero. In the case of background processes, due to the 
emission of additional particles other than $\ell^-$, $\bar{\nu}$ and \mbox{\Ds,} it is usually 
greater than zero. 
The cut \mbox{\mm $<$ 2 GeV$^2/c^4$} was applied to reduce the \Dss\ contamination.
The square recoil mass  was also used to improve the \om resolution: 
the constraint \mbox{\mm = $M^2_{\nu} (= 0 )$} 
was imposed on equation 
(\ref{eq:mm2}), which was then inverted to improve the determination of the \Bz polar angle $\theta_{\Bz}$. 
A second order equation was obtained: the resulting ambiguity was solved by choosing the solution
nearer to the previous determination. When the resolving discriminant was negative, it was forced 
to zero. This procedure improved the precision on the determination of \om both for 1992-1993 
and for 1994-1995 data samples, reducing the amount of \BtoDs~ decays outside the allowed 
kinematic range to $(4.8 \pm 0.1\%)$. The shaded area in Figure \ref{fig:omgres} shows the 
\om resolution finally obtained.

\subsection{Sample Composition}
\label{sec:sample}

A set of ${\cal{N}}_t$~= 10232 events was finally selected.
One contribution to the background was the combinatorial component, 
due to random association of a hadron 
and a lepton. Another was the resonant one, due to the association of the
lepton to a true \ps\ produced by processes different from \BtoDs. \par
  The combinatorial background was determined from the real data, by applying the previous 
selection to all candidates in the  jet containing the lepton and having the same charge as 
the lepton (wrong sign sample). A few events in this sample were in fact due to resonant processes, 
when either the lepton from \Dz~ semileptonic decay or else a fake hadron 
with the same charge as a true \ps\ was selected. Their respective amount was computed 
from the simulation as $63\pm31$ events and $76\pm38$ events. The total yield of
$139\pm 49$ events was subtracted from the wrong sign data set.
This sample was then normalised to the right sign sample by counting the events situated in 
the side band interval \mbox{0.225 GeV/$c^2 < \dm <$ 0.3 GeV/$c^2$}, 
where the fraction of events due to genuine \ps\ was negligible. 
The normalisation factor was \mbox{1.288 $\pm$ 0.012 (stat.)$\pm$ 0.021 (syst.)}  
and the corresponding number of combinatorial events in the mass interval selected for the signal
was \mbox{${\cal{N}}_{comb.}$~ = 3737 $\pm$ 70 (stat.)$\pm 75$(syst.)}. The 
systematic error consisted of three contributions. The first one
($\pm$54 events) was computed by applying the same 
procedure to the simulated data, after having removed all events containing a genuine \ps, 
in order to verify that the particles with wrong charge correlation reproduce the actual 
combinatorial background.  The difference between the right and wrong charge samples, 
after normalisation, was \mbox{44 $\pm$ 54} events. The second contribution ($\pm 49$) 
was due to the subtraction of the small amount of resonant events in the wrong charge sample
(see above). 
The residual contribution, due to leakage of \ps\ events into the side band, was negligible. \par
The total amount of \Ds was then \mbox{${\cal{N}}_{\Ds} = 6495\pm123 ({\mathrm{stat.}})\pm75 ({\mathrm{syst.}})$}.\par
The following processes contributed to the resonant background: fake leptons 
randomly associated with a \ps,
$b$ decays to a \Ds with another heavy flavour decaying semileptonically, \mbox{$b \ra \Ds X_c /\tau^- X$} 
(followed by \mbox{$X_c/ \tau^- \ra \ell^- Y$}), and production of \Dss\ in $b$ semileptonic decays.
The contribution from all these sources was determined from the simulation, 
using the most
recent measurement of the relevant branching ratios, which are reported in Table~\ref{tab:syst}. 

\par
Hadrons faking a lepton can combine with a \Ds produced either from $b\bar{b}$ or $c\bar{c}$ 
decays of the $Z$
(the contribution from gluon splitting to \Ds is negligible). Their total amount was computed
by determining independently the probability for a hadron to fake a lepton, 
known with  about $\pm 5\%$
relative precision, 
and the product of branching ratios 
\mbox{$\Br(Z \ra b\bar{b} (c\bar{c})) \times \Br(b(c) \ra \Ds$)} \cite{DELPHI:bds}.\par
The rate for the \mbox{$\Bz \ra \tau^- \bar{\nu_\tau}\Ds $} decay was obtained 
from the measurement of  the inclusive 
\mbox{\Br($b \ra \tau^- \bar{\nu_{\tau}}X_c$)} =(2.6 $\pm$ 0.4\%)
\cite{PDG}, multiplied by the probability that the charm state ($X_c$) hadronises to a \Ds.
This last number was estimated as (50$\pm$10)\% from the fraction of \Bz semileptonic decays
with a \Ds in the final state \cite{LEPBWG}.\par
The fraction of inclusive double charm decays $b \ra \Ds X_c$ was determined from charm
counting measurements as suggested in  reference \cite{HQEWWG}. 
The error on the signal included the uncertainty in the
inclusive semileptonic decay $X_c \ra \ell^- \bar{\nu_{\ell}} X$.\par
The main contribution to the resonant background is due to the intermediate production of
\Dss\ states. 
In the following it will be assumed that the \mbox{\Dss$\ra$\Ds $X$} 
decay is saturated by single particle production (namely, \mbox{$\Dss^0 \ra \Ds \pi^-$},
\mbox{$\Dss^+ \ra \Ds \pi^0$}, \mbox{$\Dss_s \ra \Ds K^0$}), a hypothesis  
consistent with the conclusions of reference \cite{DELPHI_Dss}. 
Therefore the rate for \Dss\ background production at LEP can be expressed as:
\ba
\nonumber b^{**} &=&  \Br(b\ra\ \ell^- \bar{\nu_{\ell}} \Dss) \times \Br(\Dss \ra \Ds X) \\
\nonumber        &=&  f_u \cdot \Br(B^-\ra \ell^- \bar{\nu_{\ell}} \Ds \pi^-)~+~ \\
\nonumber        & &  f_d \cdot \Br(B^0\ra \ell^- \bar{\nu_{\ell}} \Ds \pi^0)~+~ \\
\nonumber        & &  f_s \cdot \Br(B_s\ra \ell^- \bar{\nu_{\ell}} \Ds  K^0) 
\ea
where the parameters $f_u$, $f_d$, $f_s$ 
express the probability that a $b$ quark hadronises into a $B^-$, $B^0$ and $B_s$ meson
respectively (the production of \Ds from $\Lambda_b$ semileptonic decays is neglected). 
Their values are reported in Table \ref{tab:syst} as computed in reference 
\cite{LEPBWG}.
The relation $f_u = f_d$ is also imposed.\par 
By assuming that the partial semileptonic widths are the same for all $b$ hadrons,
the following relations are also derived:
\ba
\nonumber \Br(\Bz\ra \ell^- \bar{\nu_{\ell}} \Ds \pi^0) &= 
          \Br(B^-\ra \ell^- \bar{\nu_{\ell}} \Ds \pi^-) &\times 
          \frac{1}{2} \times \frac{\tau_{\Bz}}{\tau_{B^-}} \\
          \Br(\bar{B_s}\ra \ell^- \bar{\nu_{\ell}} \Ds  K^0)  &= 
          \Br(B^-\ra \ell^- \bar{\nu_{\ell}} \Ds \pi^-)&\times 
          (\frac{3}{4} \alpha ) \times \frac{\tau_{\bar{B_s}}}{\tau_{B^-}}
\label{eq:relbss}
\ea 
where the factor $\frac{1}{2}$ in the first relation accounts for isospin invariance,
the factor  \mbox{$\frac{3}{4}$} in the second one is derived from  SU(3) flavour symmetry. 
The parameter $\alpha$ = 0.75$\pm$0.25 is introduced to account for a possible 
violation of the SU(3) symmetry.
The branching ratio \mbox{$\Br(B^-\ra \ell^- \bar{\nu_{\ell}} \Ds \pi^-)$} was determined by 
the DELPHI \cite{DELPHI_Dss} and ALEPH
\cite{ALEPH_Dss} collaborations by looking for an additional 
charged pion coming from the $B^-$ decay vertex in a sample of exclusively reconstructed $\ell^- \Ds$
events. The two measurements are in good agreement and provide the average value:
\be
\Br(b\ra\ B^- \ra\ \ell^- \bar{\nu_{\ell}} \Ds \pi^-) = (4.76 \pm 0.78) \times 10^{-3}  ~~({\mathrm LEP})
\label{eq:BLEP}
\ee
The ARGUS collaboration \cite{ARGUS} has determined the fraction of \Dss\ in their
sample of \mbox{$\bar{B} \ra \Ds \ell^- X$} events  from a fit to the \mm\ spectrum. 
Using the same model assumptions as in their paper, the value
\be 
\nonumber \Br(\bar{B^0_d} \ra\ \ell^- \bar{\nu_{\ell}} \Ds \pi^0) = 
(6.2 \pm 1.9~ \mathrm{(exp.)} \pm 0.6~ \mathrm{(model)}) \times 10^{-3}
\ee
is derived, which corresponds to
\be
\Br(b\ra\ B^- \ra\ \ell^- \bar{\nu_{\ell}} \Ds \pi^-) = (5.3 \pm 1.7)\times 10^{-3}  ~~({\mathrm ARGUS})
\label{eq:BARGUS}
\ee
Equations (\ref{eq:relbss}) (\ref{eq:BLEP}) and (\ref{eq:BARGUS}) 
are finally combined to provide:
\be 
 b^{**} = (0.76 \pm 0.11 \pm 0.03 \pm 0.02)\%
\ee
where the first error is experimental, the second is due to the error on 
the B hadron production fractions (mostly $f_s$) and the third  comes from the variation
of the parameter $\alpha$ in the range 0.5 - 1.\par 
The events generated in the simulation were rescaled to the branching ratios determined previously
 and were then processed through the same analysis chain as the real data. 
This allowed the determination of the
composition of the selected \Ds sample which is reported in   
Table \ref{tab:sample}.

\bt[htb]
\bc
\begin{tabular} {|l|c|c|}
\hline
 Source                               & Amount \\ \hline
 Data                                 &  10232        \\
 Combinatorial                        &   3737$\pm$70  \\
 $\Bz \ra \Ds \tau \bar{\nu_\tau}   $ &     54$\pm$3   \\
 $b\ra \Ds X_c$                       &     56$\pm$3   \\
 fake leptons                         &    250$\pm$8   \\ 
 $b\ra \Dss \ell^- \bar{\nu_{\ell}} $ &   1469$\pm$10  \\ 
 \BtoDs                               &   4666$\pm$130 \\ \hline
\end{tabular}
\caption[]{Expected composition of the sample used for the analysis. 
Only the statistical errors are reported.}
\label{tab:sample}
\ec
\et

\boldmath
\section{Determination of \Vcb} 
\unboldmath
\label{sec:fit}

\subsection{Parametrisation of the Decay Width}
The expected number of signal events  can be expressed as a function of \om by the relation:
\ba
\nonumber d{\cal N}/d\om ~ & = &  ~ 4 ~ N_Z ~ R_b ~ f_d ~ \Br(\ps) ~\epsilon(\om)~  d\Gamma/d\om, \\
           d\Gamma/d\om  ~ &=& ~ \frac{G^2_F}{48 \pi^3 \hbar ~ \tau_{\Bz}} M^3_{\Ds} (M_{\Bz}-M_{\Ds})^2 
                            ~\sqrt{\om^2-1}~ (\om+1)^2  \\
\nonumber  &\times&  \mid V_{cb} \mid^2 ~ {\cal F}^2(\om) 
                     \left[ 1+\frac{4\om}{1+\om}\frac{1-2\om r+r^2}{(1-r)^2}\right]
\label{eq:width}
\ea
The factor 4 accounts for the fact that a \Bz can be produced in either 
hemisphere, 
and that both electrons and muons were used; $N_Z$ is the number of hadronic events, 
$R_b$ the fraction of hadronic $Z$ decays to a $b \bar{b}$ pair, $f_d$ the probability 
for a $b$ quark to hadronise into a \Bz meson, $\Br(\ps)$ the branching ratio
for the decay $\Ds \ra D^0 \pi^+$, $\tau_{\Bz}$ the \Bz lifetime
and $r$ is the ratio of meson masses, $r =  M_{\Ds}/M_{\Bz}$. The values employed for 
these parameters, as determined by independent measurements, are reported in Table \ref{tab:syst}.
 The quantity $\epsilon(\om)$, the product of the acceptance and of the reconstruction efficiency 
(which exhibits a slight dependence on \om), was determined on the basis of the tuned simulation. \par
The analytical expression of the form factor \Pfr(\om) is unknown. Because
of the small \om range allowed by
phase space, earlier analyses used a Taylor series expansion limited to second 
order:
\be
{\cal F}(\om) ~=~ {\cal F}(1) ~ (1 + \rho^2_{\cal F} (1-\om) + c (1-\om)^2 + {\cal O}(1-\om)^3)
\label{eq:Para1}
\ee
Except for \Pfr(1), theory does not predict the values of the coefficients, which must
be determined experimentally. First measurements of \Vcb\ were performed assuming a linear expansion, i.e. 
neglecting second order terms \cite{ARGUS,CLEO,ALEPH1,DELPHI:vcb}.
Basic considerations derived from the requirements of analyticity and positivity of the QCD functions describing the local currents
predict however that a positive curvature coefficient $c$ should be expected, which must be related to the
``radius'' of the heavy meson \rhf (see reference \cite{Neuold}) by the relation:
\be 
c ~=~ 0.66 {\rhf} - 0.11
\label{eq:Para2}
\ee
Results exploiting this analyticity bound have been derived by the ALEPH and OPAL 
collaborations (see reference \cite{ALEPH2,OPAL}). \par
An improved parametrisation  was subsequently proposed (see reference \cite{BGL}). It accounts for
higher order terms, so reducing to $\pm 2$\% (according to the authors) the relative uncertainty on \Vcb\ due to the 
form factor parametrisation. In this approach, the four-velocity product is first mapped onto the variable {\it z},
defined as:
\ba
\nonumber z = \frac{\sqrt{\om+1}-\sqrt{2}}{\sqrt{\om+1}+\sqrt{2}}
\ea
The form factors are then computed by continuing {\it z} in the complex plane, where it is
bound to lie within the unit circle. The form factors are then expanded as 
a power series of {\it z} while analyticity bounds and dispersion relations are employed to express 
terms up to order three as functions of the first
order coefficient. The resulting expression is rather complicated \cite{BGL}. \par
An equivalent approach was applied in reference \cite{Neunew} where the form factors 
are expressed  instead as a function of \om. In this case a novel function \Aone(\om) 
was introduced, connected to \Pfr(\om) by the following relation:
\ba
\label{eq:Para3}
 {\cal F}^2(\om)   ~&\times&~ \left[ 1+\frac{4\om}{1+\om}\frac{1-2\om r+r^2}{(1-r)^2}\right] ~=~ \\
\nonumber
 {\cal A}_1^2(\om) ~&\times&~ \left\{ 2\frac{1-2\om r+r^2}{(1-r)^2}
                                        \left[ 1+\frac{\om-1}{\om+1} R_1(\om)^2 \right] + 
                                        \left[ 1+\frac{\om-1}{1-r}(1-R_2(\om))\right]^2\right\} 
\ea 
where $R_1$(\om) and $R_2$(\om) are ratios of axial and vector form factors; their analytical expressions can be found in
reference \cite{Neunew}. The following parametrisation, depending only on a single unknown parameter \rha, was obtained for \Aone(\om):
\ba
\label{eq:APar}
 {\cal A}_1(\om) ~=~ {\cal A}_1(1) \left[ 1-8\rho^2_{{\cal A}_1}z(\om) +(53\rho^2_{{\cal A}_1}-15)z(\om)^2-
      (231\rho^2_{{\cal A}_1}-91)z(\om)^3 \right]
\ea
where the relation between {\it z} and \om was given previously.
It should be noted that in the limit \om $\ra$ 1, \Aone(\om) $\ra$ \Pfr(\om), so that \Aone(1) 
$\approx$ \Pfr(1). Experimental data were fitted using this last parametrisation and results were 
also obtained with the other forms for the sake of comparison.\par
Results using this new parametrisation have also been published by the OPAL collaboration in
\cite{OPAL2}, and presented by the CLEO collaboration at the XXX International 
Conference on High Energy Physics \cite{CLEO2}.

\boldmath
\subsection{Parametrisation of the \Dss~ Spectrum}
\unboldmath
\label{sec:dsshape}

The \Dss\ sample is composed  of four different charm orbital excitations,
two narrow resonances ($D_1$,$D^*_2$), with a measured width of about 25 MeV/$c^2$ \cite{PDG}, 
and two broad states ($D^*_1$,$D^*_0$). According to HQET, their width should be
about 200 MeV/$c^2$; the CLEO collaboration has reported preliminary evidence of the $D^*_1$ 
state with a mass of 2461$\pm$ 50 MeV and width = 290$\pm$100~MeV~\cite{CLEO_brd}.
Non-resonant \Ds$\pi$ states may also be present and contribute to the sample:
it will be  assumed in the following that their behaviour is included in that of the 
broad states.\par
The differential decay width for the decay processes \BtoDss\ has not been measured and must be 
taken from theory.
Heavy Quark Effective Theory predicts that, in the limit of infinite $b$,$c$ 
masses, the rate near
 zero recoil ( \om = 1) should be suppressed by an extra factor $(\om^2-1)$
compared to the \BtoDs\ decay rate. Several computations of the relevant 
form factors have been performed in 
this approximation (see reference \cite{Morenas} and references therein). 
However such models predict a high production rate for the $D^*_2$ states, which is 
incompatible with present
experimental information (see discussion in reference \cite{BaBar_fi_book}). \par
The effects of ${\cal O}(\Lambda_{QCD}/m_c)$ and ${\cal O}(\alpha_s)$ corrections have been
 computed in reference \cite{Leib} (referred to as LLSW model in the following).
Decay rates and differential decay widths are computed assuming two different expansions, 
(``A'' and ``B'' schemes) and the results are compared to the prediction obtained in the 
infinite mass approximation (``A$_\infty$'' and ``B$_\infty$''). A few parameters are not predicted 
by that model, but are varied within a reasonable range. When including finite $c$ mass 
corrections, the $D^*_2$ production rate decreases and is consistent with present 
experimental limits, while the \Dss\ rate near zero recoil is increased.
The form factors for the broad states are computed from those of the narrow states assuming a
non-relativistic constituent quark model with a spin-orbit independent potential. 
The model predicts that the global production rate for the broad states 
should be about equal to that of the $D_1$.
Experimental results show that narrow resonances account for about 
$(35 \pm 15)\%$  of \BtoDss\ decays \cite{LEPBWG}, in fair agreement with that prediction.
It should be noted that in the ``B$_\infty$'' scheme the rate of broad states 
is 1.65 times larger than the $D_1$ one. However, the prediction for the 
$D^*_2$ state is too high.\par
A calculation based on a relativistic quark model  (``EFG'' model) 
reduces the number of unknown parameters of the model (see reference \cite{EFG}). 
In such a case, however, broad states account for only about 25\% of
\BtoDss\ decays.\par
The following prescription, elaborated by the LEP \Vcb\ working group, was  applied to
determine \Vcb. Among all the possible expansions of the LLSW model,
only those consistent with the experimental constraint from the ratio of the $D^*_2$ to
 $D_1$ production rates were considered. This removes the A$_\infty$ and B$_\infty$
models. Each of the remaining models was then used in turn; input parameters were varied one at
a time, while keeping all the others fixed at their central value. The allowed range for each
parameter was once again determined from the $D^*_2$ over $D_1$ rate. The average of
 the two extreme \Vcb\ values so obtained was used as the measurement result, while
half their difference was considered as the systematic error. 


\subsection{Fit to the Experimental Data}
\label{sec:fitres}
Real and simulated data were collected in several \om bins. 
A minimum $\chi^2$ fit was then performed comparing the numbers of observed 
and expected events in each bin. The normalisation of the background was determined 
as explained previously. The shape of the \om distribution for
the combinatorial background was obtained from the wrong charge real data events. Simulated \Dss\ 
spectra were corrected as described in the previous section and the spectra for all the other 
background sources were taken from the simulation. 
The contribution from the signal was obtained at each step of the minimisation by 
properly weighting each generated event surviving the selection; for a given 
value of \om the 
weight was equal to the ratio between the value 
taken by the fitting function and the one of the generation function, which was 
parametrised as in equation (\ref{eq:Para1}), with \rhf$^{gen}$~=~0.8151 
and~$c^{gen}$~=~0. \par
\begin{figure}[b]
\begin {center}
\setlength{\unitlength}{1cm}
\begin{picture}(15,15)
\epsfxsize=15cm
\epsffile{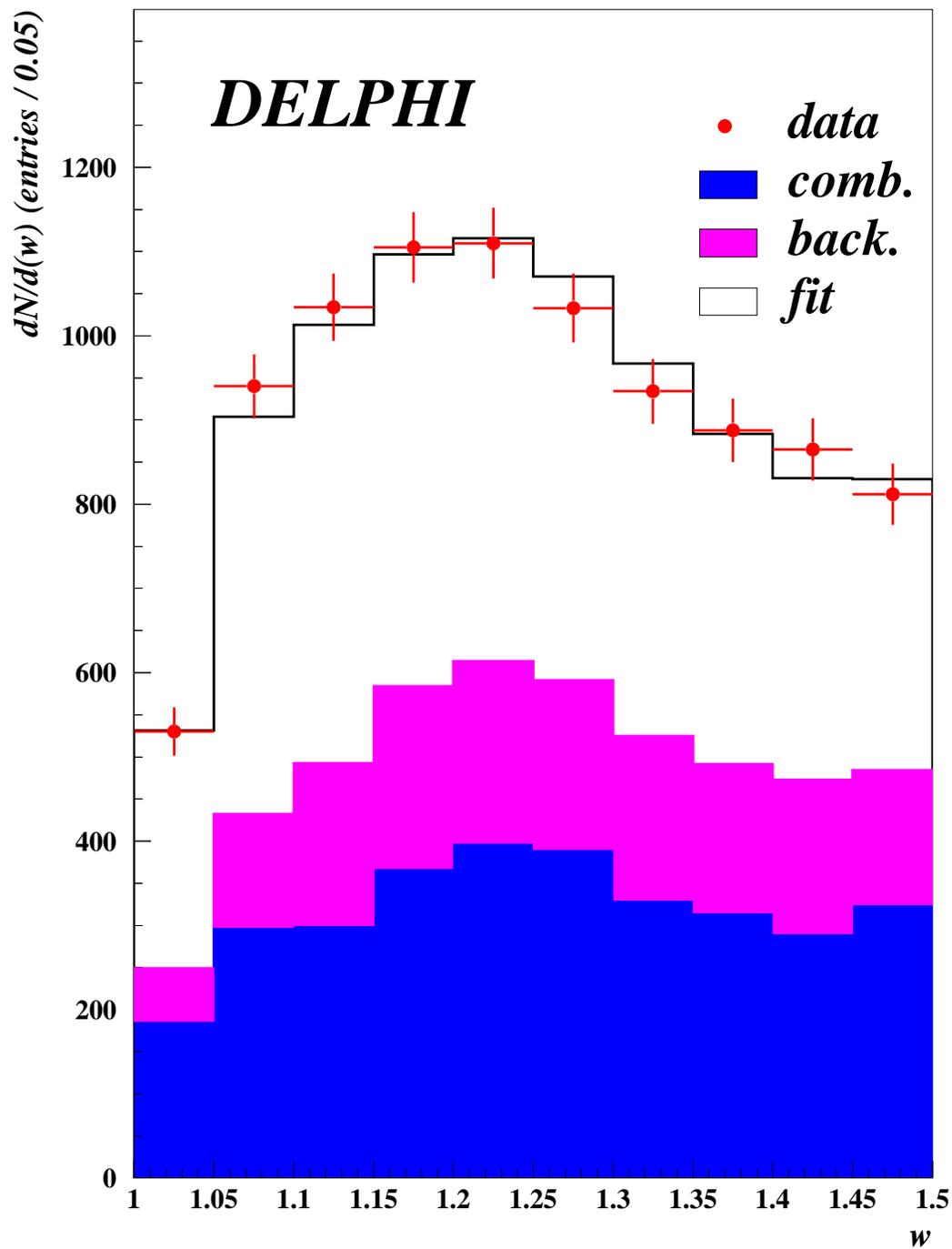}
\end{picture}
\end{center}
\caption[]{ Fit to the \om distributions. Dots with error bars: data; 
dark shaded area: combinatorial background;
light shaded area: other backgrounds, including \Dss; 
histogram: all components, the unshaded area corresponds to the decay \BtoDs. }

\label{fig:fit}
\end{figure}

Using the most recent form factor parametrisation of equation (\ref{eq:APar})
 the following results were obtained:
\ba
\nonumber {\cal A}_1(1) \mid V_{cb} \mid ~&=& ~ (35.5\pm 1.4) \times 10^{-3}\\
\nonumber  \rho^2_{{\cal A}_1}   ~&=& ~  1.34 \pm 0.14\\
\nonumber \Br({\BtoDs})         ~&=& ~ (4.70 \pm 0.12)\% 
\ea
where the last quantity was obtained by integrating the differential decay width. 
The correlation between the two fitted
parameters was 0.94.   Figure \ref{fig:fit} shows the comparison between the 
real data and the result of the fit.\par
It should be noted that the fit was performed separately on 1992-1993 and 1994-1995 data sets, and 
then the results have been averaged. 
Individual results obtained with the two data sets are in agreement, as can be seen in 
Table \ref{tab:results}.

\bt[htb]
\bc
\begin{tabular} {|l|c|c|c|c|} \hline

Fit Method (sample)             & \Vcb   \Aone(1)$\times 10^{3}$ & \rha & \rhf & $c$ \\ \hline
Eq. (\ref{eq:APar}) 92-93 & $35.8 \pm 2.5$ & $1.30\pm 0.24$     &   -  &  -  \\
Eq. (\ref{eq:APar}) 94-95 & $35.2 \pm 1.8$ & $1.40\pm 0.18$     &   -  &  -  \\
Eq. (\ref{eq:APar}) 92-95 & $35.5 \pm 1.4$ & $1.34\pm 0.14$     &   -  &  -  \\
Ref. \cite{BGL} 92-95      & $35.9 \pm 1.6$ & $-0.0009\pm0.021$  &   -  &  -  \\       
Eq. (\ref{eq:Para1})~92-95 & $35.8 \pm 1.4$ &   -                & $1.22\pm 0.14$   &  0.66\rhf$-$0.11 \\
Eq. (\ref{eq:Para1})~92-95 & $36.9 \pm 1.9$ &   -                & $1.59\pm 0.41$   &  $1.4\pm 0.9$  \\
Linear               92-95 & $34.6 \pm 1.3$ &   -                & $0.90\pm 0.10$   &  0  \\
Linear    94 only          & $36.4 \pm 1.5$ &   -                & $0.84\pm 0.12$   &  0  \\
Ref. \cite{DELPHI:vcb}     & $35.9 \pm 2.2$ &   -                & $0.74\pm 0.20$   &  0  \\ \hline
\end{tabular}
\caption[]{Results of different fits to the experimental data. Results in the fourth line are 
obtained assuming the form factor representation of reference \cite{BGL}, where the expansion is
performed directly on {\it z}. For this reason the first order Taylor 
coefficient (\rha)  cannot be
compared directly to the corresponding one of reference \cite{Neunew} in the first three lines.}

\label{tab:results}
\ec
\et

The same table also contains the results obtained when using the other parametrisations of form factors
discussed previously. In detail, the Taylor expansion of equation 
(\ref{eq:Para1}) was employed, by assuming a
linear expansion ($c$=0), by imposing the constraint of equation 
(\ref{eq:Para2}) for the curvature $c$ or by 
fitting \rhf~ and $c$ as independent free parameters. In this last case, the correlation coefficients with 
\mbox{ \Vcb \Pfr(1)} were 0.82 and 0.71 respectively, the mutual correlation 
was 0.97. The last two entries in the table
are reported to show the consistency with the published result of reference \cite{DELPHI:vcb}.
The measurement was performed by using the same data sample (1994) and the same model assumptions 
for the signal and the background as in that previous publication, 
but applying the new data selection and \om reconstruction.

\section{Systematic Uncertainties}
\label{cha:sys}

The individual sources of systematic errors are reported in Table \ref{tab:syst}
and are described in detail below.
Uncertainties in the  overall normalisation, 
in the knowledge of the selection efficiency and of the composition of the sample, 
including the  modelling of the background, and about the agreement
between the experimental  and the simulated resolution may affect the results. They 
were all considered as sources of systematic error. 

The fit was performed several times, by  varying in turn all the parameters which determine the normalisation 
(see equation (\ref{eq:width})) within their allowed range. The corresponding variations of the
measured quantities were added quadratically to the systematic error.

The efficiency depends on the detector performance in track reconstruction, lepton identification and secondary
vertex reconstruction.
Tracks from charged particles may be lost because of cracks in the tracking device, or because of hard scattering
of the particle by the detector frames. Electrons and low momentum \ps\ are more sensitive to this last effect.
A conservative error of $\pm 1\%$ per track was assigned, based on studies of the detector material (performed using
electrons from photon conversion) and of the TPC cracks.\par
The actual efficiency for lepton identification was measured exploiting samples of real data
tagged independently. Muons from $\tau$ decays and from the process 
$\gamma\gamma \ra \mu^+\mu^-$ were used  to explore
all the relevant kinematic range. The values of the experimental and predicted efficiencies were consistent 
within $\pm 2\%$.
Electrons from photon conversion and from the radiative Bhabha process were also used. 
Compared to the simulation, a relative
efficiency of  \mbox{$(94 \pm 2)\%$} was found, where the error is due to the 
systematic difference between the two samples.
This ratio does not depend on the particle momentum.\par
To provide an accurate description of the algorithm employed for vertex reconstruction, 
the simulation was tuned following
the procedure of reference \cite{Borisov}, developed for the precise measurement of $R_b$. 
The efficiency was then determined by comparing in the real data and in the simulation the fraction of 
vertices  reconstructed in a sample with high momentum leptons.
The ratio between the efficiency found in the real data to the one found in the simulation 
was 1.01$\pm$0.01. The average number of charged particles forming the inclusive vertex 
in the simulation was slightly larger than in the real data. This was attributed to a small 
loss in the efficiency to  assign the charged particle tracks to the secondary vertex. 
The ratio of the  vertex reconstruction efficiencies was estimated to be $0.99 \pm 0.01$. 

Because of the cuts on the lepton momentum and decay length significance \Sgn~ 
(see Section \ref{sec:selection}), the efficiency  depends on the average fraction of the beam 
energy actually carried by the \Bz meson, $< x_E >$.
Events were generated assuming the Peterson fragmentation function \cite{Pete}, tuned so as to 
reproduce the measured value of \mbox{$< x_E > ~=~ 0.702\pm0.008$} \cite{HQEWWG}; they were then 
reweighted in the fit in  order to allow for a variation of
$\pm$ 0.008 in $< x_E >$, and the consequent change of the fitted parameters 
was propagated into the errors. \par
Model dependent uncertainties may be introduced by the kinematic cuts on \ptl~ and \mm~ as well. 
They were determined following the iterative procedure applied in reference~\cite{DELPHI:vcb}.
The simulated spectrum was corrected to the measured values 
and the efficiency computed again. The efficiency varied by about $\pm$ 1\%. 
This was taken as the systematic error, 
and no further iteration was performed.

Each source of background was changed by its error as given in Table \ref{tab:sample} and the 
variation of the results was propagated into the error. \par 
The reconstruction efficiency and the \om resolution depend on the multiplicity of charged 
particles in \Dz\ decays, improving
with higher values. It should be noted that zero-prong \Dz\ decays are also collected, albeit with 
smaller efficiency and with worse \om resolution, in all the cases in which a secondary vertex 
can be formed by the lepton and the \ps\ alone.
The simulation was tuned to the results of the MARK III measurement \cite{MARKII}.
The relative fractions of events  with 0, 2, 4, $>$4 charged prongs were then varied 
within their errors to compute the systematic error. \par
The fraction of $K^0$ produced per \Dz\ decay was varied as well, to account for possible loss of 
efficiency and degradation of resolution due to the presence of $K^0_L$. \par
The systematic error due to the knowledge of the \Dss\ spectra was determined following the
prescriptions of the LEP \Vcb Working Group, as explained in detail in section \ref{sec:dsshape}.

All quantities relevant to the determination of the \om resolution were studied.
Agreement was found between the distributions of the \Ds energy in the real data and in the simulation; 
the angular resolution on the \Ds direction was checked by comparing the relative angle between 
the \ps\ and the \Dz\ directions and again very good agreement was found.\par
The estimate of the \Bz energy depends on the hermeticity of the detector. 
To verify that cracks were properly simulated,
a sample of $b$ enriched events was provided by $b$-tagging one hemisphere and analysing the other 
(unbiased). Only hemispheres without identified leptons were considered, in order to avoid possible 
distorsions due to the presence of a neutrino. The procedure was applied to the experimental data and
 to the simulation, and the visible energy was compared in the two cases. Depending on the year, the 
energy seen in the real data (about 37 GeV) exceeded 
the one predicted by about 200-400 MeV. The main source of discrepancy was attributed to the tracking. 
Due to the smallness of the effect, no further investigation was performed but two different correction
procedures were followed: either the visible energy in the simulation was 
increased by the relevant amount,
or else a correction was computed depending on the fraction of charged energy seen in the event.
The systematic error was chosen as the maximum difference between the result so obtained and the one 
without fine tuning.\par
The \Bz angular resolution was compared in the real data and in the simulation by inspecting the angle
between the \Bz and the \Ds$\ell^-$ directions. The RMS widths of the two distributions were
identical within errors (42.5$\pm$0.6 and 42.6$\pm$ 0.6 mrad respectively). The systematic error 
on \Vcb\ was computed by repeating the fit without the \mm\ constraint. 
The \om resolution depends on the precision of the vertex reconstruction, which improves
when more charged tracks form the vertex. Events in the simulation
were rescaled to correct for the small discrepancy in the vertex multiplicity discussed 
previously and the fit was repeated. The difference was negligible.\par
The systematic error induced by the fitting procedure was determined 
by varying the number and the size of the bins, and by removing the (few) events outside the 
physically allowed \om region. The effect of the other cuts applied in the analysis was 
checked by varying them in the ranges :
\bi
\item the \dm~ cut  between 0.15 and 0.20 GeV/$c^2$,
\item the \ptl~ cut between 0.8 and 1.25 GeV/$c$,
\item the \mm~  cut between 0. and 5. GeV$^2$/$c^4$ 
\item the \Sgn~ cut between 2.5 and 6.5
\ei

The efficiency and purity of the sample vary by more than \mbox{50\%}
in these ranges, and most of the induced variations are compatible with statistical fluctuations.
They were conservatively assumed as systematic errors.
As a  further check, the analysis was 
performed separately for electrons and muons. Excellent agreement was found.
All the errors were added in quadrature to determine the final systematic uncertainty.
\bt[htb]
\bc
\begin{tabular} {|l|c|c|c|c|}
\hline
 Parameter       & Value       &     $\frac{\D(\Vcb)}{\Vcb} (\%) $   &    \D(\rha)    
                 & $\frac{\D(\Br(\BtoDs)}{\Br(\BtoDs)} (\%) $  \\ \hline
 $R_b$                                      & $(21.66\pm 0.07)\%$ \cite{HQEWWG} &    0.25  &   -  &  0.5  \\
 {$\Br$}(\Ds $\ra$ \Dz$\pi^+$)           & $(67.7\pm0.5)\%$  \cite{PDG}      &    0.43  &  -   &  0.9 \\
 {$\Br$}(\Bz $\ra$ \Ds$\tau^-\bar{\nu_{\tau}}$) & $(1.3\pm0.3)\% $   \cite{HQEWWG}  &    0.23  &  -   &  0.19 \\
 {$\Br$}($b   \ra  {\Ds}X$)              & $(23.1\pm1.3)\%$    \cite{DELPHI:bds} &   0.05  &  -   &  0.19 \\
 {$\Br$}($c   \ra  {\Ds}X$)              & $(24.0\pm1.3)\%$    \cite{DELPHI:bds} &   0.13  &  -   &    -  \\  
 {$\Br$}($b   \ra  {\Ds}X_c$,            &  &  & & \\ 
 \hspace{0.75cm}$X_c \ra \ell^-\bar{\nu_{\ell}}X$) 
                                            &$(0.87^{+0.23}_{-0.19})\%$\cite{HQEWWG}&   0.15  & 0.01 &  0.30 \\
 {$\Br$}(\Dz $\ra  \ell^+ \nu_{\ell} X$)        & $(6.75\pm0.29)\%$   \cite{PDG}        &   0.10  &  -   &   -   \\
 {$\Br$}(\Dz $\ra  K^0 X$)               & $(42 \pm 5  )\%$    \cite{PDG}        &   0.20  & 0.01 &  0.60 \\  
            \Dz\ decay mult.                &   see \cite{MARKII}                   &   0.10  &  -   &  0.20 \\  
 {$\Br$}(\BtoDss) & & & & \\
 $\times ~\Br(\Dss\ra\Ds X)$          & $(0.77\pm0.11)\%$ &   1.09  & 0.05  &  4.40 \\
 \Dss\ model                           &                   &   5.10  & 0.20  &  0.20 \\
 $<x_E>$               & $(70.2\pm0.08)\%$ \cite{HQEWWG}  &   0.99  &   -   &  1.90 \\ \hline
 $f_d$ = {$\Br$}($b   \ra \Bz$)    &$(40.5\pm 1.1)\%$ \cite{LEPBWG}     &   1.90  & 0.02 &  2.70 \\
 $f_s$ = {$\Br$}($b   \ra B_s$)    &$( 9.5\pm 1.3)\%$ \cite{LEPBWG}     &   0.08  &   -  &  0.20 \\
 $f_\Lambda$ = {$\Br$}($b\ra\Lambda_b$)&$(9.5\pm 1.9)\%$  \cite{LEPBWG}  &   0.13  &   -  &  0.41 \\
 $\tau_{\Bz}$                        & $(1.55\pm 0.03)$ps \cite{PDG}     &   1.22  &   -  &  0.74 \\  
 $\tau_{B^+}$                        & $(1.65\pm 0.03)$ps \cite{PDG}     &   0.03  &   -  &  0.20 \\ 
 $\tau_{B_s}$                        & $(1.49\pm 0.06)$ps \cite{PDG}     &   0.03  &   -  &  0.20 \\
  \hline
 Tracking                            &                   &   1.00  &  -    &  2.00 \\
 Secondary Vertex                    &                   &   0.50  & 0.02  &  1.80 \\             
 $\ell$ eff. \& bgd.                 &                   &   0.70  &   -   &  1.50 \\
Combinatorial                        &                   &   0.52  & 0.02  &  1.87 \\
E$_\nu$ tuning                       &                   &   0.21  & 0.01  &  0.19 \\ 
fit                                  &                   &   0.23  & 0.02  &  0.20 \\
\Sgn\ vertex                         &     2.5-6.5       &$^{+0.20}_{-0.70}$ & $^{+0.01}_{-0.06}$ & $^{+0.50}_{-0.10}$\\
$P_t$ lepton                         & 0.8-1.25 (GeV/$c$)  &$^{+0.02}_{-.90}$ &  -0.03  & 1.30  \\
\dm                                 &0.15-0.20(GeV/$c^2$)&$^{+2.10}_{-1.90}$& $^{+0.12}_{-0.05}$ & 3.40   \\
\mm                              &0.0-5.0 (GeV$^2$/$c^4)$&$^{+1.30}_{-0.70}$& $^{+0.05}_{-0.02}$ & $^{+2.40}_{-0.30}$\\
\om resolution                           & no \mm\ constraint  &   -2.10          & -0.07              &   -0.50   \\ \hline
Total Systematic         &  & $^{+6.4}_{-6.8}$ & $^{+0.24}_{-0.22}$& $^{+7.6}_{-6.9}$      \\ \hline  

\end{tabular}
\caption[]{Contributions to the systematic uncertainties. 
The values used for the parameters relevant to this analysis
are reported in the second column. Errors for \Vcb\ and \Br(\BtoDs) 
are relative and given in \%; the errors for \rha are absolute.}
\label{tab:syst}
\ec
\et

\section{Extraction of the form factor}

The result of the previous section  was obtained in the framework 
of a specific model. It is in principle possible to extract the differential decay width, 
$d\Gamma/d\om $, 
from the experimental data. To cope with the non-negligible smearing due to the 
experimental resolution, an unfolding procedure was applied \cite{Blob,Zech}. 
With this same technique the Isgur Wise function, 
the universal form factor expected in the framework  of the HEQT, was also extracted.\par 
The simulated events which survived the selection were first grouped in ten bins, 
according to the value of $\om_{gen}$ at generation. Because of the finite 
experimental resolution, 
events lying inside a given bin in
$\om_{gen}$ populated several bins in the reconstructed $\om_{rec}$ 
distribution. For each 
$\om_{gen}$ bin, a corresponding $\om_{rec}$ histogram was obtained. To overconstrain the
fit, the new histograms consisted of twelve bins. The linear combination of these ten histograms
was fitted to the real data distribution. The ten parameters of the fit determined the 
normalisation coefficients for each simulation histogram. 
The unfolded differential decay width was finally 
obtained by binning the simulated events according to the value of $\om_{gen}$ and scaling the 
resulting histograms with the fitted parameters.   \par
\begin{figure}[b]
\begin{center}
\setlength{\unitlength}{1cm}
\begin{picture}(15,15)
\epsfxsize=15cm
\epsffile{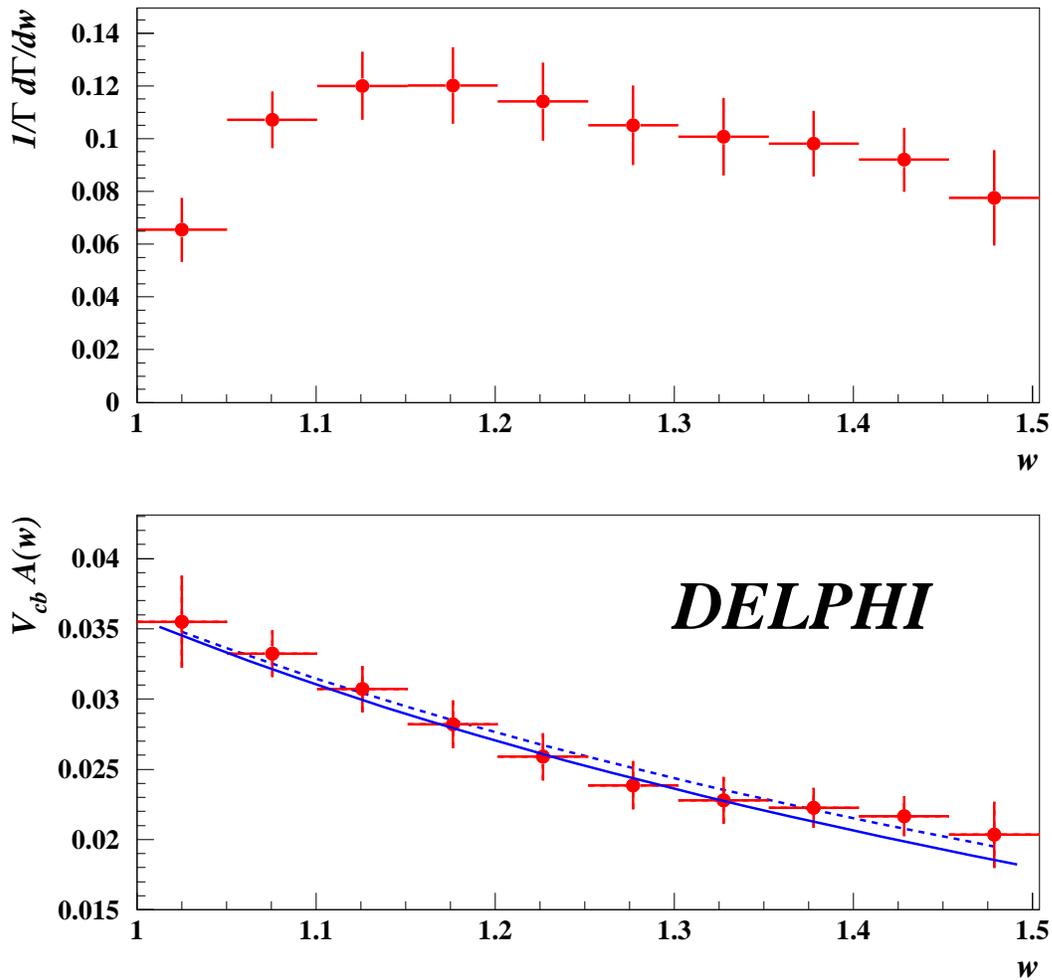}
\end{picture}
\end{center}
\caption[]{Unfolded distributions in the real data. 
Upper plot: differential decay width. 
Lower plot: decay form factor as in equation (\ref{eq:Para3}). The dotted 
line shows the results of a fit to the histograms, neglecting bin-to-bin 
correlations. The continuous line shows the result obtained when including the
statistical correlations among the bins.}

\label{fig:unfold}
\end{figure}

To avoid spurious bin-to-bin oscillations, typical of such an unfolding method, a 
regularisation term was added to the $\chi^2$, which is proportional
to the second derivative of the unfolded results:
\begin{equation}
\nonumber \chi^2_{reg} ~~~~=~~~~ \tau \cdot \Sigma_{i=2}^{n-1}|(f_{i+1}-f_{i})-(f_{i}-f_{i-1})|^2 
              ~~~~\propto~~~~ \tau \cdot \int |f''(x)|^2 dx
\end{equation}
The regularisation parameter $\tau$ is in principle arbitrary. Too small values lead to 
oscillating solutions, whereas large values produce flat solutions with small errors 
and strong positive correlations among parameters.
Several fits were performed with $\tau$ ranging from 0.01 to 1.0. To test the method, the 
unfolded distributions were fitted with the function of equation (\ref{eq:APar}) neglecting 
bin-to-bin correlations. The values obtained for
\Aone $\cdot$ \Vcb\ and \rha were always well compatible with those given in section 
\ref{sec:fit}, but their errors depend on the choice of $\tau$ (lower values leading to higher 
errors). Choosing $\tau$=0.20, the same errors as 
the ones of section \ref{sec:fit} were obtained.
The corresponding unfolded spectra are presented in Figures 
\ref{fig:unfold}(a,b); 
the dashed curve overlayed shows the result obtained when fitting neglecting bin-to-bin correlation.
To remove the sensitivity of the errors to the choice of $\tau$, fits were finally performed 
properly accounting  for bin-to-bin correlations: they are represented by the continuous line
in Figure \ref{fig:unfold}(b). The result was:
\ba
\nonumber \Aone \cdot \Vcb &=& (36.1 \pm 1.4) \times 10^{-3} \\
\nonumber \rha &=& 1.38 \pm 0.15
\ea
independent of the choice of $\tau$. The small difference from the values presented in section 
\ref{sec:fitres} is interpreted as the systematic error due to the unfolding procedure. 
The unfolded data and their error matrix are presented in Table 
\ref{tab:unfold}.

\bt[htb] 
\bc
\begin{tabular} {l|cccccccccc} \hline
$\om$-1       & .025 & .075 & .125 & .175 & .225 & .275 & .325 & .375 & .425 & .477 \\ \hline
$1/\Gamma \cdot d\Gamma/d\om$ (\%) 
              & 6.5  & 10.7 & 12.0 & 12.0 & 11.4 & 10.5 & 10.0 & 9.8  & 9.2  & 7.8  \\ \hline
              & .140 &&&&&&&&& \\
              & .053 & .108 & &&&&&&& \\
              &-.055 & .042 & .160 &&&&&&& \\
              &-.086 &-.035 & .093 & .199 &&&&&& \\
              &-.058 &-.068 &-.015 & .101 & .205 &&&&& \\
              &-.013 &-.058 &-.075 &-.021 & .109 & .215 &&&& \\
              & .016 &-.026 &-.077 &-.080 &-.002 & .128 & .210 &&& \\
              & .026 & .003 &-.041 &-.075 &-.066 & .003 & .099 & .147 && \\
              & .021 & .024 & .006 &-.035 &-.082 &-.096 &-.053 & .043 & .134 & \\
              & .014 & .037 & .043 & .003 &-.081 &-.160 &-.170 &-.065 & .145 &0.308 \\
\end{tabular}
\caption[]{The unfolded differential decay width. First line: \om bin centre. Second line: 
partial decay width in the corresponding bin, divided by the total width, expressed in percent.
Other lines: corresponding error matrix expressed in permill. All the values were obtained with 
the regularisation constant $\tau = 0.20$.}
\label{tab:unfold}
\ec\et


\section{Conclusions}

A sample of about 5000 \BtoDs~ decays was obtained by means of the method of the inclusive 
\ps\ tagging, originally developed at LEP by the DELPHI collaboration. The use of the large 
data set, and the excellent detector performance
allowed the precise measurement of the product \Vcb$ \cdot \Aone(1)$~ and of the \Bz 
``radius'' \rha, following the most recent parametrisation of the Isgur-Wise function 
proposed in reference \cite{Neunew}:
\ba
\nonumber \mid V_{cb} \mid \cdot {\cal A}_1(1) &=& 
~(35.5 \pm 1.4({\mathrm stat.}) ^{+2.3}_{-2.4} ({\mathrm syst.}))\times10^{-3} \\
\nonumber \rha                                 &=& 
~  1.34 \pm 0.14({\mathrm stat.}) ^{+.24}_{-.22}({\mathrm syst.}) \\
\nonumber \Br(\BtoDs) &=& 
~ (4.70 \pm 0.13 ^{+0.36}_{-0.31})\%
\ea
Using the value \Aone(1) $\approx$ \Pfr(1) = 0.91 $\pm$0.03, the following value of 
\Vcb\ is  obtained:
\ba
\nonumber \mid V_{cb} \mid ~=~(39.0 \pm 1.5 ({\mathrm stat.}) ^{+2.5}_{-2.6}
 ({\mathrm syst.~exp.}) \pm 1.3 ({\mathrm syst.~th}))\times 10^{-3}
\ea
These results agree with the present world average (see reference \cite {PDG}).
They supersede  the previous DELPHI measurement of reference \cite{DELPHI:vcb}.\\

\subsection*{Acknowledgements}
\vskip 3 mm
 We wish to thank D. Ebert, R. Faustov, B. Grinstein, Z. Ligeti, M.
 Neubert and M. Wise
 for useful discussions, and R. Lebed for kindly providing the code describing his
 model.
 We are greatly indebted to our technical 
collaborators, to the members of the CERN-SL Division for the excellent 
performance of the LEP collider, and to the funding agencies for their
support in building and operating the DELPHI detector.\\
We acknowledge in particular the support of \\
Austrian Federal Ministry of Education, Science and Culture,
GZ 616.364/2-III/2a/98, \\
FNRS--FWO, Flanders Institute to encourage scientific and technological 
research in the industry (IWT), Belgium,  \\
FINEP, CNPq, CAPES, FUJB and FAPERJ, Brazil, \\
Czech Ministry of Industry and Trade, GA CR 202/96/0450 and GA AVCR A1010521,\\
Commission of the European Communities (DG XII), \\
Direction des Sciences de la Mati$\grave{\mbox{\rm e}}$re, CEA, France, \\
Bundesministerium f$\ddot{\mbox{\rm u}}$r Bildung, Wissenschaft, Forschung 
und Technologie, Germany,\\
General Secretariat for Research and Technology, Greece, \\
National Science Foundation (NWO) and Foundation for Research on Matter (FOM),
The Netherlands, \\
Norwegian Research Council,  \\
State Committee for Scientific Research, Poland, 2P03B06015, 2P03B11116 and
SPUB/P03/DZ3/99, \\
JNICT--Junta Nacional de Investiga\c{c}\~{a}o Cient\'{\i}fica 
e Tecnol$\acute{\mbox{\rm o}}$gica, Portugal, \\
Vedecka grantova agentura MS SR, Slovakia, Nr. 95/5195/134, \\
Ministry of Science and Technology of the Republic of Slovenia, \\
CICYT, Spain, AEN96--1661 and AEN96-1681,  \\
The Swedish Natural Science Research Council,      \\
Particle Physics and Astronomy Research Council, UK, \\
Department of Energy, USA, DE--FG02--94ER40817. 

\newpage 

\newpage 

\newpage 

\begin{figure}[b]
\begin {center}
\setlength{\unitlength}{1cm}
\begin{picture}(15,15)
\epsfxsize=15cm
\epsffile{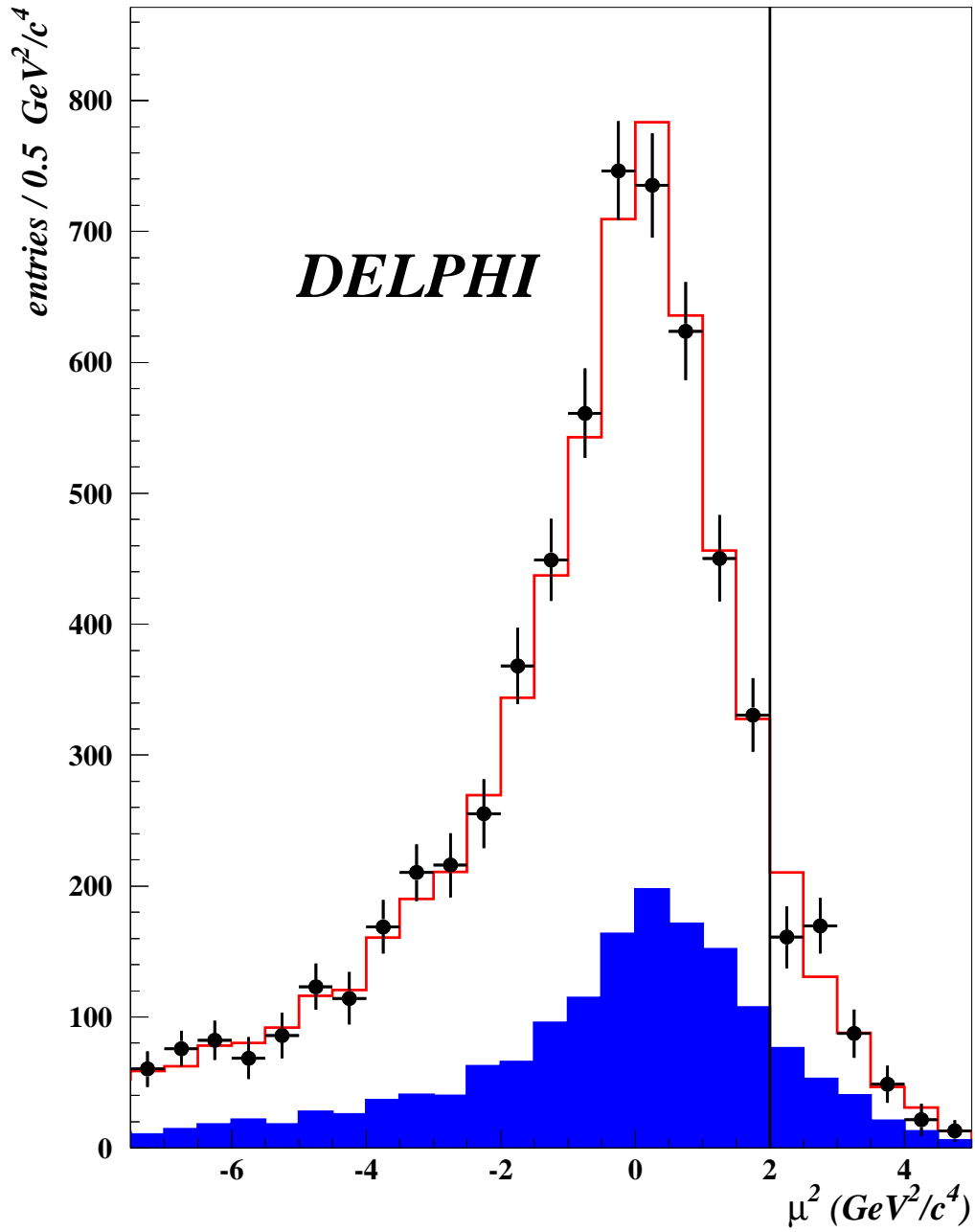}
\end{picture}
\end{center}
\caption[]{ Squared missing mass distributions. 
Real data (dots with error bars)
 after subtraction of all the background apart from \Dss\ 
are compared to the sum of the \Ds and \Dss
contributions as predicted by the simulation; 
the dark area represents the \Dss. The vertical line shows the position of the
cut.}

\label{fig:mm2}
\end{figure}

\newpage 

\newpage 

\end{document}